\documentclass[aip,jcp,reprint]{revtex4-1}
\usepackage{amsmath}
\usepackage{amsfonts}
\usepackage{url}
\usepackage{bm}
\usepackage{graphics}
\usepackage{paralist}

\usepackage{subfig}
\usepackage{graphicx}

\usepackage[bookmarks=false,colorlinks=true,linkcolor=black,urlcolor=black,citecolor=black]{hyperref}

\newcommand{\ket}[1]{|#1\rangle}
\newcommand{\bra}[1]{\langle #1|}

\newcommand{\e}[1]{{\mbox{e}}^{#1}}

\begin{document}

\title{Assessment of Multireference Approaches to Explicitly Correlated Full Configuration Interaction Quantum Monte Carlo}

\author{J. A. F. Kersten}
\email{jennifer.kersten@cantab.net}
\affiliation{University Chemical Laboratory, Lensfield Road, Cambridge, CB2 1EW, U.K.}
\affiliation{Max Planck Institute for Solid State Research, Heisenbergstra{\ss}e 1, 70569 Stuttgart, Germany}
\author{George H. Booth}
\email{george.booth@kcl.ac.uk}
\affiliation{Department of Physics, King's College London, Strand, London WC2R 2LS, U.K.}
\author{Ali Alavi}
\email{a.alavi@fkf.mpg.de}
\affiliation{University Chemical Laboratory, Lensfield Road, Cambridge, CB2 1EW, U.K.}
\affiliation{Max Planck Institute for Solid State Research, Heisenbergstra{\ss}e 1, 70569 Stuttgart, Germany}

\begin{abstract}

The Full Configuration Interaction Quantum Monte Carlo (FCIQMC) method has proved able to provide near-exact solutions to the
electronic Schr\"odinger equation within a finite orbital basis set, without relying on an expansion about a reference state. 
However, a drawback to the approach is that being based on an expansion of Slater determinants, the FCIQMC method suffers from a basis set incompleteness error 
that decays very slowly with the size of the employed single particle basis. 
The FCIQMC results obtained in a small basis set can be improved significantly with
explicitly correlated techniques. Here, we present a study that assesses and compares two contrasting `universal' explicitly correlated approaches that fit into the FCIQMC framework; 
the $[2]_{R12}$ method of Valeev {\em et al.}, and the explicitly correlated canonical transcorrelation approach of Yanai {\em et al}. The former is
an {\em a posteriori} internally-contracted perturbative approach, while the latter transforms the Hamiltonian prior to the FCIQMC simulation. These comparisons
are made across the 55 molecules of the G1 standard set. We found that both methods consistently reduce the basis set incompleteness, for accurate atomization energies
in small basis sets, reducing the error from 28~mE$_{\text h}$ to 3-4~mE$_{\text h}$. While many of the conclusions hold in general for any 
combination of multireference approaches with these methodologies, we also consider FCIQMC-specific advantages of each approach.
\end{abstract}

\maketitle
\section{Introduction}

The vast majority of the properties of everyday molecular and solid state systems can be predicted from the solution of the non-relativistic, time-independent 
electronic Schr\"odinger equation. Static properties, which depend only on the ground state solution, cover important experimental quantities such as ionization potentials
or atomization energies, required for heats of formation. However, the complexity of the fully coupled equations results in its exact solution being out of reach for all
but the smallest systems to date.
The Full Configuration Interaction Quantum Monte Carlo (FCIQMC) method\cite{Booth_Alavi_2009}, is a diagonalization-free, non-perturbative approach which 
projects a stochastic sampling of the wavefunction towards the exact ground state
solution of the Schr\"odinger equation within a given orbital basis (OBS). This is achieved via a stochastic propagation of a population of walkers within the full space of Slater determinants. The dynamics are devised such that the long-time average over the discrete walker population on any determinant approaches that of the FCI coefficient.
For several systems\cite{Booth_Alavi_2010, Booth_Alavi_2011} it has been shown that the FCIQMC method is capable of providing the exact FCI energies within 
systematically improvable errorbars of $\mathcal{O}[10^{-4}-10^{-5}]E_h$, with only a fraction of the computational costs of an equivalent FCI calculation. 
Thus this method can be applied to tackle difficult systems that require an accurate and non-perturbative description of the correlation.
In the past, the FCIQMC method has be employed to calculate the energies of a diverse range of systems, including atoms and molecules\cite{FCIQMC_N2,
FCIQMC_ETHENE,FCIQMC_diatomics,Thomas_2015}, the Hubbard model\cite{FCIQMC_Hubbard}, the uniform
electron gas\cite{Shepherd_Alavi_2012_1}, and solid state
systems\cite{FCIQMC_solidstates}.
Since its first implementation, the efficiency and abilities of the FCIQMC method have been steadily improved. 
The initiator approximation\cite{Booth_Alavi_2011, Cleland_Alavi_2010, Cleland_Alavi_2011}, real walkers, and semi-stochastic sampling\cite{semi_stochastic, Blunt_SS} have
decreased the computational effort of the method considerably, and new functionalities include the calculation of excited
states\cite{excited_states_QMC,Blunt_2015,Chan_2012,Nagase2011,Ten-no_MSQMC,blunt2015excited},
the calculation of accurate two-particle reduced density matrices using a replica sampling technique\cite{DensityMatrix_Blunt,RDM_Overy}, and the use of these density matrices within a
complete active space self-consistent field (CASSCF) approach\cite{Thomas_CASSCF, Manni_2016}.

While an FCIQMC calculation is generally far less computational expensive than a conventional FCI
calculation, FCIQMC calculations are still limited to relatively small orbital basis sets, because the number of Slater determinants scales binomially with their size.
This results in basis set incompleteness error (BSIE) in computed energies and properties, compounded by the fact that this error decreases slowly with
respect to increasing basis set size\cite{L3_convergence1, L3_convergence2,Morgan_1992,Shepherd_Alavi_2012_2}. The origin of this error is the result of the 
inability of Slater determinants to model the electron-electron cusps in the wavefunction\cite{Slater.31.333,Hylleraas_1929,Pack_cusp1}. 
As described by Kato's cusp condition\cite{Kato_1957}, the cusp regions are defined where the inter-electronic distance $r_{12}=|{\bf r}_1 - {\bf r}_2|$ goes to zero. Away from this point, 
the wavefunction depends linearly on $r_{12}$. Such behaviour is very difficult to obtain from a superposition of smooth Slater determinants which are built from
one-electron functions that are centred at the nuclei. Thus very large basis sets are required to provide the flexibility for an accurate description of these
cusp regions. 

An alternative to extrapolating this basis set incompleteness through often expensive calculations in large basis sets is given by explicitly correlated methods\cite{R12_rev1,
Kong_Valeev_2012, rev_Tew}, which can reduce the BSIE efficiently with often little additional
cost compared to the original calculation. 
The general aim of these approaches is to augment the ansatz for the wavefunction with explicitly correlated geminals which contain terms that 
are linear in the inter-electronic distance $r_{12}$ at the cusp positions.
Solutions for a large number of technical challenges from the inclusion of these geminals have been addressed in the literature over the last few 
decades\cite{R12_Kutzelnigg,Kutzelnigg_MP2r12, Kutzelnigg_standardA, Noga_CCR12, Noga_CCR12_2,ABS,Valeev2004190,DF1, DF2,Tenno_2007}.
These most notably include the avoidance of up to four-electron integrals, which have been addressed with efficient resolution of the identity techniques, and an optimized
Slater-type geminal form, which provides a more efficient description of the wider correlation hole about each cusp position\cite{Tenno_2000_1, Tenno_2000_2, Tenno_2004, Tew_geminal,r12f12,Tenno_MRF12}.
This work has led to modern `F12' methods which are reliable, efficient and robust.

Whilst earlier efforts focused on the combination of F12 methodology with single-reference quantum chemical techniques, there has been increasing emphasis on
adapting the technology for use with multi-reference methods, including multireference configuration interaction\cite{Gdanitz_MRCI-R12,Shiozaki_MRCIF12} and 
multireference perturbation theories\cite{Tenno_MRF12,Werner_MRF12}, within an consistent, internally contracted framework. 
This has been an important advance, since even though multireference methods are required in the presence of stronger, static correlations in low-spin open-shell or transition state systems for example, 
these correlations are present {\em in addition} to the cusp-dominated dynamic correlation required for quantitative accuracy. These already challenging systems therefore also inherit 
all of the same slow basis set convergence problems of single reference systems.
This is further exacerbated by the generally far higher cost with respect to basis set size for these systems, and therefore their combination with explicitly correlated
techniques is a significant improvement.
However, a drawback of many of the multireference approaches to date is that they are tailored and embedded within each parent multireference method, and therefore new equations have to be
derived and code implemented, for each new multireference method. 

This issue is rectified by the introduction two `universal' explicitly correlated multireference techniques as the topic of the 
current study, which can be simply applied to almost all electronic structure methods. Their combination with the FCIQMC method provides a powerful approach to multireference quantum
chemistry, and their overall accuracy, as well as suitability for use with FCIQMC will be assessed. These approaches are the $[2]_{R12}$ method of Torheyden, Kong and Valeev\cite{Torheyden_R12,
Kong_Valeev_2011}, which is briefly described in section~\ref{sec:R12}, and the canonically transcorrelated approach of Yanai {\it et al.}\cite{Yanai_Shiozaki_2012, CT_Yanai}, 
described in section~\ref{sec:CT_theory}.
These provide contrasting approaches to the challenge of a universal multireference F12 method, with the former providing an internally contracted, perturbative coupling of the 
geminals to a given multireference wavefunction, while the latter uses an initial guess density in order to transcorrelate the Hamiltionian, effectively removing the cusp features
from its solution, prior to the multireference treatment. These approaches differ substantially in the way that the geminals are coupled to the multireference wavefunction, and in
whether the multireference wavefunction can relax due to their presence, with the former adhering to a more traditional `diagonalize-then-perturb' approach, while the latter
is constructed on a `perturb-then-diagonalize' philosophy.
Whilst the infinite basis limit is exact for both approaches, away from this limit, they are not
expected to behave the same. Previous applications of these approaches to FCIQMC have been shown to improve the results significantly via the reduction in BSIE, however
these studies were limited to very few systems and did not allow for a direct comparison between the two different methods\cite{Booth_Tew_2012, FCIQMC_CT}.
In this paper, the two explicitly correlated FCIQMC approaches are assessed in an in-depth study using the 55 molecules of the G1 standard set\cite{G1_set_1,G1_set_2}, and by using
this large sample size it is possible to draw more general conclusions about the success and efficiency of both methods.

\section{Methods}
\subsection{FCIQMC}
In the Full Configuration Interaction Quantum Monte Carlo (FCIQMC) method, the wave function $\ket{\Psi}$ is expanded in a basis set of all $N$-electron Slater determinants
$\{\ket{D_{I}}\}$,
\begin{align}
\ket{\Psi} = \sum_{I}\ C_{I} \ket{D_{I}}.
\label{FCIQMC_ansatz}
\end{align}
The coefficients $C_{I}$ are coarse-grained and stochastically sampled using a population of walkers, where each walker is defined by its sign, weight, and the
Slater determinant it is associated with. The walkers then evolve according to a set of coupled differential
equations derived from the imaginary time Schr\"odinger equation, such that
 the long-time average of the signed weight of walkers $n_{I}$ on a Slater determinant $D_{I}$ is proportional to $C_{I}$: 
\begin{align}
n_{J}(\tau+ \Delta \tau)=n_{J}(\tau)
-\Delta\tau \sum_{I}    (H_{JI}-E_S \delta_{IJ}) \ n_{I}(\tau) ,
\label{FCIQMC_walker}
\end{align}
 where $H_{JI}$ are the matrix elements of the Hamiltonian operator, $E_S$ an energy offset, and $\Delta \tau$ a small 
interval in imaginary time.

Following Equation (\ref{FCIQMC_walker}), three consecutive steps are performed every time step $\Delta\tau$.
\begin{itemize}
   \item \textbf{Spawning}: Each occupied determinant $D_{I}$ spawns new walkers onto $\chi_I$ randomly chosen connected determinants $D_{J\neq I}$, with
\begin{align}
\chi_I = 
\left\{
\begin{array}{ll}
&\lceil | n_I | \rceil \; \; \textrm{with probability} \; | n_I |- \lfloor | n_I | \rfloor, \\\
&\lfloor |  n_I |  \rfloor \; \; \textrm{otherwise},
\end{array} \right. 
\end{align}
where $\lfloor x \rfloor$ denotes the largest integer that is not
greater than $x$, and $\lceil x \rceil$ denotes the smallest integer that is not
smaller than $x$.
The sign of the newly spawned walkers on a determinant $D_{J}$ is the same as parent determinant if $H_{J, I}<0$ and opposite otherwise, and their
weight is given by
\begin{align}
p_{\text{s}}(J|I)=\frac{\delta \tau |H_{J, I}|}{p_{\text{gen}}(J|I) },
\end{align}
where $p_{\text{gen}}(J|I)$ denotes the probability of choosing the determinant $D_{J}$. More details on the specifics of this spawning step can be found in Ref.~\onlinecite{RDM_Overy}.
   \item \textbf{Death/Cloning}: The weight of walkers $n_I$ on each determinant $D_{I}$ is reduced by a stochastically realized amount given by
\begin{align}
p_{\text{d}} =\delta \tau \left(H_{I,I} - E_{S}\right) n_I.
\end{align}
   \item \textbf{Annihilation}: At the end of each time step, the list of newly spawned walkers is merged with the list of the old walkers (for algorithmic details of 
       how to implement this efficiently, see Ref.~\onlinecite{Booth_algorithm}). Walkers that are
assigned to the same determinant but have an opposite sign annihilate each other, so that all walkers on each occupied determinant have the same sign at the end of each iteration.
\end{itemize}
 A typical FCIQMC calculation starts with one walker on the Hartree--Fock (HF) determinant and a fixed energy offset $E_S= E_{\text{HF}}$. Due to the fact that the 
 lowest eigenvalue of the Hamiltonian must be below $E_{\text{HF}}$, 
 the population of walkers grows exponentially until a specified number of walkers is reached. Then $E_S$ is varied smoothly, so that the walker
population becomes approximately constant. The ground state energy can be determined from the long-time average of the energy offset $E_S$ and from the
long-time average of the projected energy:
\begin{align}
\langle E_p\rangle= E_{\text{HF}} +   \frac{\langle \sum_{J \neq 0}\  H_{0,j}\ n_{ J}\rangle}{\langle
n_{0}\rangle},
\end{align}
where ${\langle .. \rangle}$ denotes an imaginary-time average, and $n_{0}$ denotes the weight of walkers on the reference wavefunction. Often this is taken to be the largest
weighted single Slater determinant, but can also be an arbitrary linear combination of Slater determinants\cite{Blunt_SS}. Once the wavefunction has converged, and is sampling the
desired solution, it is also possible to accumulate the two-body reduced density matrix, which is sampled at the same time as the spawning steps above. As this is a non-linear
function of a random variable (the sampled wavefunction), serious systematic errors can result in this sampling. This issue is remedied via the introduction of a second `replica' sampling
of the wavefunction, which samples randomly from the same distribution of the ground state wavefunction, but is uncorrelated to the first. This allows for an unbiased sampling
of the reduced density matrices from which many properties of the wavefunction derive. Whilst the sampling of the density matrices is now unbiased, there are computational overheads
with its computation. The sampling of the second replica approximately doubles the computational overhead, both in terms of memory and processor time, while additional (non-distributed) memory
is required to store the full two-body density matrix (currently in non-sparse form). More details on the sampling of the density matrices, as well as more details on the 
FCIQMC algorithm can be found in Refs.~\onlinecite{RDM_Overy,Booth_algorithm,Booth_Alavi_2009,Booth_Alavi_2010, Booth_Alavi_2011}.

\subsection{The $[2]_{R12}$ Approach}\label{sec:R12}
The spin-free  $[2]_{R12}$ approach developed by Valeev and co-workers can be employed \textit{a posteriori} to 
reduce the basis set incompleteness error of an arbitrary multi-reference calculation\cite{Torheyden_R12,Kong_Valeev_2011,Luke_MPQC}.
The $[2]_{R12}$ correction to the energy is evaluated in an internally contracted, second-order perturbative fashion via the 
Hylleraas functional. The first order wave function is expanded as explicitly correlated geminal replacements, constrained to be orthogonal, two-electron excitations to a
multi-configurational reference wave function $\ket{0}$. 
In contrast to the MRMP2-F12 ansatz developed by Ten-no\cite{Tenno_MRF12}, the explicitly correlated $[2]_{R12}$ wave function also
includes semi-internal excitations into geminal functions, and so the full first-order wavefunction can be written as
\begin{align}
 \ket{\psi ^{(1)}}   =&  \hat{\Omega}^{(1)} \ket{0}, \label{Omega}\\
  =& \frac{1}{2} t_{rs}^{pq}
\left(
r_{\alpha^{\prime}\beta^{\prime}}^{rs}
\hat{E}^{\alpha^{\prime}\beta^{\prime}} _{pq}
+
2 r^{rs}_{\alpha^{\prime} x} \hat{E}^{\alpha^{\prime} x} _{pq}
\right)
\ket{0} \nonumber\\
&-\frac{1}{2} t_{rs}^{pq}
\left(
2 r^{rs}_{\alpha^{\prime} k} \left(\gamma^{(-1)} \right)_j^i \Gamma_{pq}^{jk} \hat{E}^{\alpha^l
{\prime}
} _{i}
\right)
\ket{0},
\end{align}
where $r$ are the matrix elements of the correlation factor, $\gamma$ and $\Gamma$ are the spin-free
one-body and two-body reduced density matrices (RDM) of the reference wave function, and
$\hat{E}$ are the spin-free excitation operators. 
The notation of the orbitals belonging to the different parts of the orbital space is shown in Table
\ref{Tab:obital_space}, and all equations are written using the Einstein summation convention, i.e.
repeated indices are implicitly summed. 

\begin{table}[t]
\centering 
\begin{tabular}{l l}
\hline
\hline 
Orbital space
& 
Notation
\\
 [0.5ex]
\hline
Correlated orbitals 
&
$p, q, r, s, t, u , v, w$
\\
Occupied orbitals
&
$i, j, k$\\
Orbital basis sets (OBS)
&
$x, y, z$\\
Complementary auxiliary basis sets (CABS)\ \ \ 
&
$\alpha^{\prime}, \beta^{\prime}$\\
Complete virtual space
&
$\alpha, \beta$\\
Formally complete basis set (CBS)
&
$\kappa, \lambda$\\
\hline
\end{tabular}
 \caption{\small Notation of the orbital space.}
\label{Tab:obital_space}
\end{table}
The geminal coefficients $t_{rs}^{pq}$ are fixed according to the SP ansatz so
that they satisfy the cusp conditions\cite{Tenno_2004},
\begin{align}
t^{pq}_{rs} = \frac{3}{8} \delta_r^p \delta_s^q + \frac{1}{8} \delta_r^q \delta_s^p .
\end{align}
The second order Hylleraas function yields:
\begin{align}
 \mathcal{H}^{(2)}  & =
\bra{\Psi_1} \left[  \hat{F}_N , \hat{\Omega}^{(1)} \right] + \hat{\Omega}^{(1)}  \hat{F}_N \ket{0}
+ 2 \bra{0} \hat{H}^{(1)} \hat{\Omega}^{(1)} \ket{0} ,
\end{align}
where $\hat{F}_N$ is the normal ordered spin-averaged Fock operator.\\
The $\bra{\Psi_1}\hat{\Omega}^{(1)}  \hat{F}_N \ket{0}$ term contains 4-body RDM terms, but its calculation can be avoided with the assumption that
 the generalized Brillouin condition is valid:
\begin{align}
\mathcal{H}^{(2)}  & \approx
\bra{\Psi_1} \left[  \hat{F}_N , \hat{\Omega}^{(1)} \right] \ket{0} +2 \bra{0} \hat{H}^{(1)} 
\hat{\Omega}^{(1)} \ket{0} ,
\end{align}
The matrix elements are evaluated using the expanded Wick's theorem, and the resulting expressions depend on the 1-body, 2-body, and 3-body reduced density
matrices. The 3-body RDM terms are approximated with 1-body and 2-body terms by neglecting the irreducible 3-body cumulant in the
generalized normal ordering of Kutzelnigg-Mukherjee\cite{Mukherjee,Kutzelnigg_NO,DCFT_Kutzelnigg}.
Additionally, all terms that are quadratic in the 2-body RDM cumulants are discarded. Also, the screening approximation is 
employed and all terms in which a geminal matrix element 
and another geminal matrix element or a Coulomb matrix element 
that are connected over a 2-body cumulant vanish as well. 
The correlation factor, 
\begin{align}
    f_{12} =  -\frac{1}{\gamma}\e{-\gamma r_{12}},      \label{eqn:Geminal}
\end{align}
is fitted using 6 Gaussian functions to simplify the evaluation of the integrals.
Further details on these approximations, along with the final working equations of the theory
can be seen in Refs.~\onlinecite{Torheyden_R12,Kong_Valeev_2011}.

In our work, the 2-body RDMs are sampled with the FCIQMC method along with the correlation energy using the stand-alone 
code \texttt{NECI}\footnote{The FCIQMC code can be obtained from {\tt https://github.com/ghb24/NECI\_STABLE.git}}. 
Whilst we have also implemented our own program for the $[2]_{R12}$ corrections\cite{Booth_Tew_2012}, in this work we use the interface to
the $[2]_{R12}$ implementation within \texttt{MPQC}\cite{mpqc1, mpqc2}, reading in the sampled density matrices and orbital information 
using an interface developed by Roskop {\it et al.}\cite{Luke_MPQC}.

\subsection{Canonical Transcorrelation Theory}\label{sec:CT_theory}

`Transcorrelated' methods, where the Hamiltonian operator is transformed by a Jastrow-style operator which compensates part of the correlated physics,
has a long history in electronic structure, starting with Hirschfelder in 1963\cite{hirschfelder1963}, and extended by Boys and 
Handy\cite{Boys_Handy_1969_1,Boys_Handy_1969_2, Boys_Handy_1969_3}. These approaches used a similarity transformation of the Hamiltonian operator
which rendered the resulting operator non-hermitian, and numerically problematic. These ideas have been further developed by other authors to
avoid many of the original shortcomings\cite{Tenno_2000_1, Tenno_2000_2,Umezawa_Tsuneyuki_2003, Umezawa_Tsuneyuki_2004,Luo_2010, Luo_2011, Luo_2012}.
More recently, building on the work of White\cite{White_2002}, Chan, Yanai and coworkers developed a related approach, but where the operator
(trans)correlating the Hamiltonian is unitary. This yields a now hermitian effective Hamiltonian, and is called Canonical Transformation (CT) 
theory\cite{Neuscamman_2010,Yanai_2006,CT_Yanai,Yanai_2007_2,Neuscamman_09,Neuscamman_10,Neuscamman_10_2,Yanai_12,Watson_16}.

In the explicitly correlated version of canonical transformation theory developed by Yanai and Shiozaki\cite{Yanai_Shiozaki_2012},
the parameters for the unitary transformation operator, $e^{\hat{A}}$ with $-\hat{A} = \hat{A}^\dagger$, are obtained from the projection of a set of strongly orthogonal 
explicitly correlated geminal functions. These are found as
\begin{align}
& \hat{A} = \frac{1}{2} \bar{R}_{ij}^{\alpha \beta}  \left(
\hat{E}_{ij}^{\alpha \beta}  -
\hat{E}^{ij}_{\alpha \beta} \right), \\
& \bar{R}_{ij}^{\alpha \beta}
= \frac{3}{8} \bra{\alpha \beta} \hat{Q}_{12} f_{12} \ket{ij} 
+
\frac{1}{8} \bra{\alpha \beta} \hat{Q}_{12} f_{12} \ket{ji} ,
\\
& \hat{Q}_{12} = (1-\hat{O}_1) ( 1-\hat{O}_2) -\hat{V}_1 \hat{V}_2 ,
\end{align}
where the one-electron operators $\hat{O}$ and $\hat{V}$ project onto the occupied orbitals and the
virtual orbitals of the OBS respectively, the SP ansatz has been used to fix the geminal coefficients, and the same Slater-type geminal form is used as in
Eq.~\ref{eqn:Geminal}.
The effective Hamiltonian is derived as
\begin{align}
\hat{H}_{\text{TC}} &= e^{\hat{A}^\dagger} \hat{H} e^{\hat{A}} , \\
&= \hat{H} + \left[ \hat{H}, \hat{A} \right] + \frac{1}{2!} \left[\left[ \hat{H}, \hat{A} \right],
\hat{A}\right] + \dots, 
\end{align}
where $\hat{H}_{\text{TC}}$ is hermitian, and the transcorrelated eigenvalue
problem can be solved with the traditional post-HF methods that are built on the variational
principle. 

In order to simplify the calculation, several approximations are made for the transcorrelated
Hamiltonian.
Firstly, the Baker-Campbell-Hausdorff expansion is truncated after the second order.
Secondly, all commutators are approximated with one- and two-body operators, denoted with $\left[
\ldots \right]_{1, 2}$. Similar to the
decomposition and approximation of the 3-particle density matrix using the 
 Kutzelnigg-Mukherjee formalism, the higher order
excitation operators are replaced with an approximate decomposition into one- and two-body
terms. In addition, in the second order term, the Hamiltonian
is
replaced by the Fock operator $\hat{F}$:
\begin{align}
\hat{H}_{\text{TC}} 
&\approx \hat{H} + \left[ \hat{H}, \hat{A} \right]_{1, 2} + \frac{1}{2!} \left[\left[ \hat{F}, \hat{A}
\right], \hat{A}\right]_{1, 2} .
\end{align}
The resulting terms for the matrix elements of the transcorrelated Hamiltonian are similar to the
intermediates emerging in standard F12 calculations and the many-electron integrals are solved in a similar manner using
standard approximation C,\cite{SA_C} and RI insertions, resulting in an effective Hamiltonian which is only two-body, and 
can therefore be treated by standard electronic structure methods.
%

The electron density which is used in order to define the initial geminal functions must be supplied prior to the calculation of the effective 
Hamiltonian, technically resulting in a loss of its state-universal characteristics. 
In this work, we use a trial density obtained from a prior CASSCF calculation, in order to allow the geminals to be constructed in the presence of
any static correlation. 
The matrix elements of the transcorrelated Hamiltonian are then computed with the stand-alone code
\texttt{ORZ}, using the F12 integral engine of Shiozaki\cite{Shiozaki2009CPL,Shiozaki2011JUQ,Shiozaki2009program}, 
constructing the geminal functions within the same active space that was chosen for the CASSCF calculation. 
Since the resulting transcorrelated Hamiltonian is hermitian and of two-particle form, the FCIQMC method is then employed to find the lowest 
energy eigenvalue.

\section{Results}

In order to assess the quality of the two explicitly correlated FCIQMC approaches, we considered the calculation of total energies and atomization energies
of the 55 molecules across the G1 standard set\cite{G1_set_1, G1_set_2}. FCIQMC used in conjunction with the $[2]_{R12}$ method is denoted FCIQMC-R12 and the 
canonical transcorrelation approach with CT-FCIQMC.
The G1 molecules are relatively small, single-reference dominated systems that have been extensively studied in the past\cite{G1_1, G1_2,G1_DFT, G1_DFT_1, G1_3, G1_4}.
All geometries were fully optimized with \texttt{GAMESS} at the MP2/aug-cc-pVQZ level.
All other calculations were performed with frozen cores to reduce the computational effort. 

The FCIQMC calculations were performed with the initiator approximation\cite{Cleland_Alavi_2011}, the semi-stochastic method\cite{semi_stochastic,Blunt_SS} and 
the replica sampling where the RDM is required\cite{RDM_Overy,DensityMatrix_Blunt}. The size of the deterministic space for the semi-stochastic 
adaptation was chosen to be one tenth of the size of the
initiator space unless this was prohibited by memory limitations ($\mathcal{O}[10^4 - 10^5]$ determinants). For the closed shell systems, time-reversal symmetry was also
employed\cite{Booth_Alavi_2011}. 
We obtained the correlation energies and statistical errors from Flyvbjerg-Petersen blocking analyses of the projected energies\cite{Blocking}.
For all cases, the statistical errors were found to be very small ($\mathcal{O}[10^{-5}]\, \text{E}_{\text{h}}$), and so statistical error bars are generally not visible in the following figures.
The leading error is hence the initiator error, which can be systematically improved by increasing the
total number of walkers. For the molecules of the G1 standard set, the total number of walkers
was chosen individually, depending on the convergence of the system, between 50 million and 1 billion walkers.
While the chosen number of walkers was sufficient to ensure that the initiator error in the energies of the smaller systems is negligible, 
it cannot be guaranteed that the energies of the larger molecules are entirely free from initiator error. 
Quantifying the remaining initiator error is difficult, but we estimate it to be $\mathcal{O}[10^{-4}]\text{E}_{\text{h}}$ for the total energy of the largest systems. This 
is certainly small enough not to affect any resulting conclusions or qualitative trends in the results.
For the explicitly correlated corrections, we set the parameter $\gamma$ to $1\, \text{a}_{\text{0}}^{-1}$, and used the aug-cc-pVDZ-OPTRI
basis sets for the construction of the CABS where possible. 
 

Along with incompleteness in the description of two-electron part of the wavefunction, there is also a basis set incompleteness in the one-electron description
of the 
wavefunction\cite{molpro_CCSD_F12,rev_Tew,Booth_Tew_2012}, which is excluded from the construction of these universal F12 corrections.
While the presence of strong correlations can change the magnitude of this one-electron incompleteness, it is generally much smaller than the two-electron
incompleteness, and so we assume here that it is well represented by the incompleteness in the uncorrelated Hartree--Fock energy. To estimate the complete basis set (CBS) limit of
this one-electron energy, we extrapolated the HF energies to the CBS limit with an exponential dependence\cite{HF_extra}, as
\begin{align}
E^{\text{HF}}_{\text{CBS}}  \approx \frac{E_X ^{\text{HF}}- b E_{X-1}^{\text{HF}}}{1-b},
\label{HF_extra}
\end{align}
with
\begin{align}
b= \frac{E_X ^{\text{HF}}- E_{X-1} ^{\text{HF}}}{E_{X-1} ^{\text{HF}}- E_{X-2} ^{\text{HF}}},
\end{align}
where $X$ is the cardinal number of the basis set and calculations in aug-cc-pVTZ, aug-cc-pVQZ, and aug-cc-pV5Z basis sets were used. We note that it is possible for other
methods to be used to estimate the one-electron basis set incompleteness from correlated wavefunctions\cite{Kong_2010}. 

\subsection{Total Energies}

\begin{figure*}[ht!!]
\begin{center} 
\includegraphics[clip=true, height=9cm, width=0.8\textwidth]{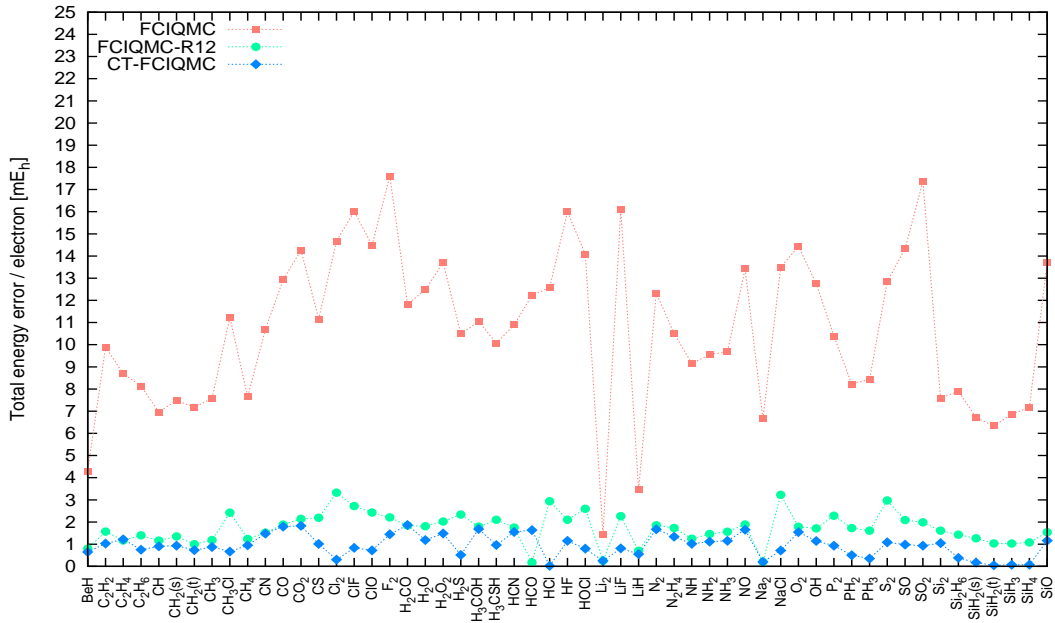} 
\caption{Absolute errors of the total energies per correlated electron for the G1 set of molecules, calculated with the FCIQMC method, and the FCIQMC-R12 and CT-FCIQMC methods
    (with one-electron BSIE corrections) in aug-cc-pVDZ basis sets.
\label{fig_tot_energy}} 
\end{center} 
\end{figure*}

The molecules of the G1 standard set have a generally single-reference character so that CCSD(T) calculations are capable of achieving high accuracy results. It is likely then
that these results will be fairer comparison than experimentally derived results for the total energy, due to uncertainties in the experimental procedure for this quantity.
Hence, the assumed exact energies in the CBS limit that we used as a benchmark were obtained from CCSD(T)-F12b calculations in
aug-cc-pV5Z basis sets performed with \texttt{MOLPRO}\cite{MOLPRO, molpro_CCSD_F12, molpro_CCSD_F12_2}, which we then further correct for static correlation effects. 
The static correlation corrections for the different molecules were approximated with the individual error of a CCSD(T) calculation in an aug-cc-pVDZ basis set
compared to the near-exact FCIQMC result in the same basis set. Since the majority of the static component of correlation can be captured 
in a small orbital space (hence the success of the CASSCF-derived approaches to strongly correlated systems), it is reasonable to assume that this 
correction will account for most of the deficiencies in the CCSD(T) results.
As expected, the static corrections are small for those single-reference systems: The absolute mean error is 0.1~$\text{mE}_{\text{h}}$ per correlated
electron, and the maximum absolute total error of 3.0~$\text{mE}_{\text{h}}$ (0.4~$\text{mE}_{\text{h}}$ per correlated electron) was found for the Si$_2$
molecule.

In Fig.~\ref{fig_tot_energy} we compare the FCIQMC results in aug-cc-pVDZ basis sets with the results from FCIQMC-R12 and CT-FCIQMC calculations with added
one-electron BSIE corrections for all 55 molecules of the G1 standard set. It can be seen that both explicitly correlated methods reduce the 
absolute error per correlated electron significantly for all molecules. The effect of the correction is stronger for the
systems that show a larger BSIE, such as SO$_2$ and F$_2$.
Furthermore, for nearly all molecules, the CT-FCIQMC method achieves slightly lower errors than the
FCIQMC-R12 method. On average, the CT method and the $[2]_{R12}$ method reduce the absolute error in the total energy per correlated electron in an aug-cc-pVDZ basis 
from 10.8~$\text{mE}_{\text{h}}$ to 1.0~$\text{mE}_{\text{h}}$ and 1.8 $~\text{mE}_{\text{h}}$ respectively.

\subsection{The Atomization Energies}

\begin{figure*}[ht!]
\begin{center} 
\includegraphics[clip=true, height=9cm, width=0.8\textwidth]{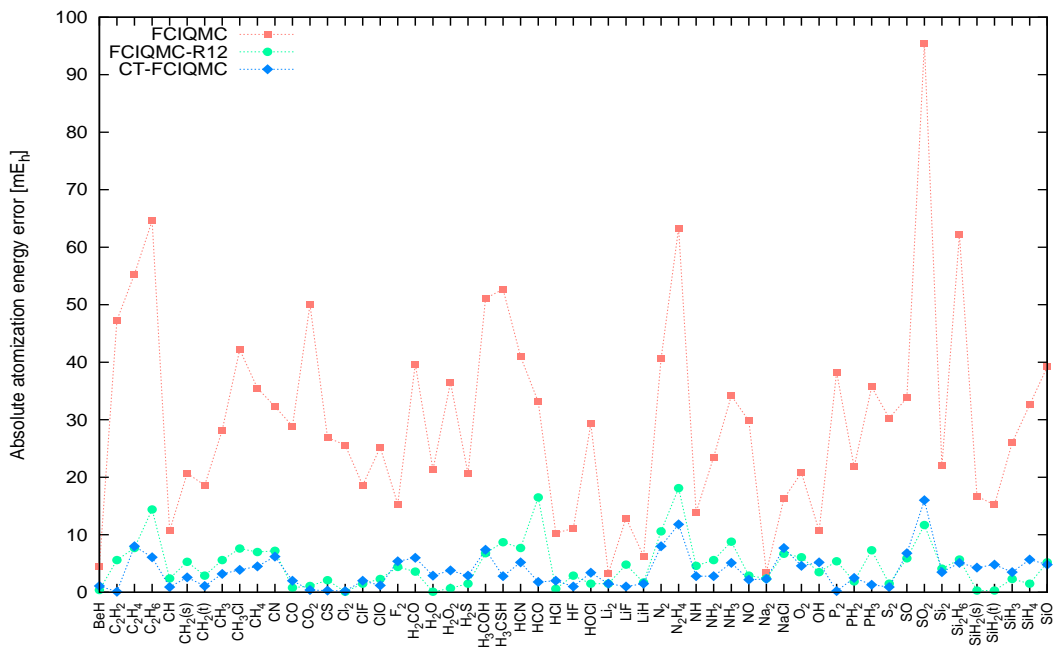} 
\caption{Absolute errors of the atomization energies for the G1 set of molecules, calculated with the FCIQMC method, and the FCIQMC-R12 and CT-FCIQMC
    methods (with one-electron BSIE corrections) in aug-cc-pVDZ basis sets.
\label{fig_atomization}} 
\end{center} 
\end{figure*}

We also compared the performance of the explicitly correlated corrections for the atomization energies. As benchmark results, we use experimental
data\cite{G1_b31, G1_b32, G1_b33, G1_b34, G1_b35, G1_b36, G1_b37, G1_b38, G1_b39, G1_b40,G1_b42, G1_b73,G1_b80, G1_b81, Dattani2015, LeRoy2009} corrected for
zero-point
vibrations, spin-orbit coupling, core-valence effects, and scalar relativistic effects\cite{feller_benchmark}. 
In Fig.~\ref{fig_atomization} the absolute error is shown for the atomization energy of each molecule. Again, we can see that the corrections for the basis set incompleteness 
error improve the FCIQMC results drastically. For this case of relative energies, the difference between the CT and the $[2]_{R12}$ method is small.

\begin{figure}[h]
\begin{center} 
\includegraphics[clip=true, width=0.46\textwidth]{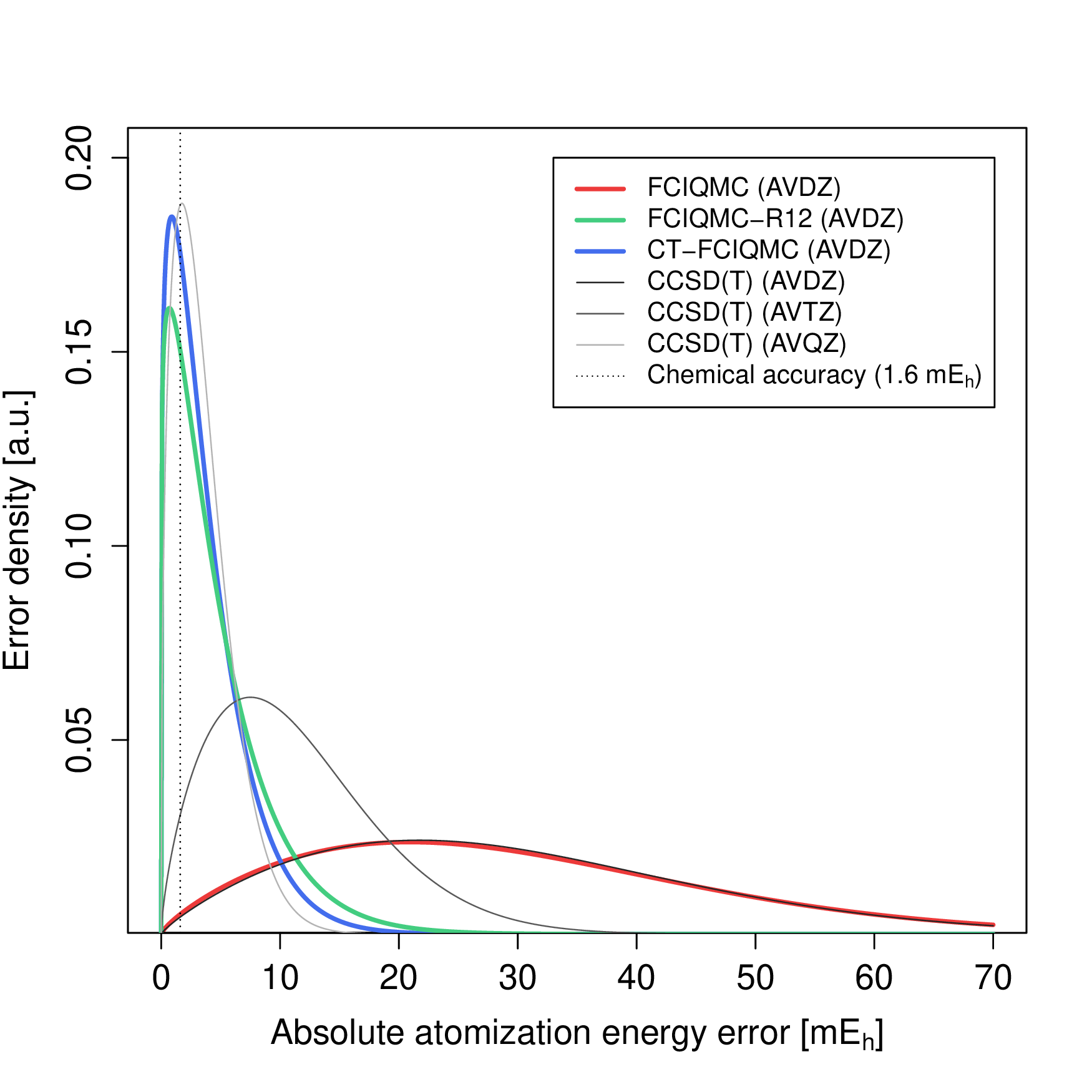} 
\caption{\small  Distribution of the absolute errors in the atomization energies of the G1 set, fitted to Weibull distributions.
\label{fig_atomization_dist}} 
\end{center} 
\end{figure}

To better analyse and compare the distribution of errors between the different methods, Fig.~\ref{fig_atomization_dist} shows the density of the errors in the 
atomization energies achieved with different methods fitted to a Weibull distribution\cite{Weibull1, Weibull2}.
We can clearly see that the
median of the error distribution across the set is reduced from 28.2$~\text{mE}_{\text{h}}$ to about 2.9$~\text{mE}_{\text{h}}$ with the CT correction and to 
4.1$\text{mE}_{\text{h}}$ with the $[2]_{R12}$ correction. 
Thus, the FCIQMC-R12 and the FCIQMC-CT methods are capable of producing relatively accurate atomization energies in aug-cc-pVDZ basis sets (though not quite
chemical accuracy), 
with far greater reliability than without the explicit correlation, indicated by the smaller spread and median in the results.
It is notable that the spread and the median in these errors is slightly larger for the FCIQMC-R12 method, compared to the CT method, suggesting a marginally
less reliable reduction in the error with the approach.

Fig.~\ref{fig_atomization_dist} also shows the atomization error distributions obtained with CCSD(T) calculations without explicit correlation in increasingly
large basis sets. 
The CCSD(T) results are mostly very accurate within the given basis set for the molecules of the G1 standard set, as shown by the similarity of the CCSD(T) and FCIQMC distributions
for the aug-cc-pVDZ basis.
However, as has been noted elsewhere, the explicitly correlated calculations in
aug-cc-pVDZ basis sets achieve a similar accuracy as CCSD(T) calculations in aug-cc-pVQZ basis sets, already without including the additional static correlation component of the
electronic structure which is obtained via the FCIQMC. Thus it can be concluded that the R12 and the CT method both effectively
decrease the BSIE to a level that is normally achieved with basis sets that are two cardinal number higher. 

\subsection{The Initiator Error}
\begin{figure}[h]
\begin{center} 
\includegraphics[clip=true, width=0.46\textwidth]{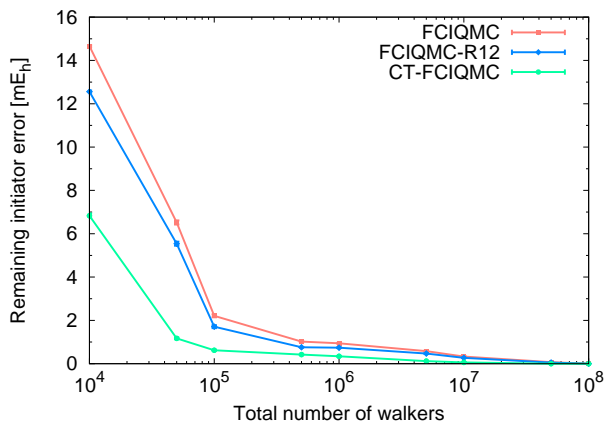} 
\caption{\small  Decay of the initiator error with respect to the total number of walkers for the FCIQMC, the FCIQMC-R12, and the CT-FCIQMC approaches for the N$_2$ molecule in an aug-cc-pVDZ basis.
\label{fig_initiator}} 
\end{center} 
\end{figure}

The initiator approximation reduces the computational effort of an FCIQMC calculation significantly, but introduces an initiator error into the sampled
energies. This error can be reduced systematically by increasing the total numbers of walkers, and the nature of this convergence is illustrated for
the FCIQMC, FCIQMC-R12, and CT-FCIQMC methods for the N$_2$ molecule in Fig.~\ref{fig_initiator}.
It can be seen that the CT-FCIQMC energy converges noticeably faster than the FCIQMC energy with respect to the total number of walkers. To reach convergence
within 1~mE$_\text{h}$, the CT-FCIQMC calculation requires less than $10^5$ walkers while the conventional FCIQMC calculation needs more than $5\times10^5$
walkers to achieve the same level of convergence. 
This indicates that removing the short-ranged part of the Coulomb hole from the Hamiltonian improves the convergence of the initiator error, and fewer walkers and thus less
computational effort is required to determine the ground state energy of the transcorrelated Hamiltonian accurately. This is despite having the same number of degrees of freedom and
overall Slater determinants as the other calculations.
The $[2]_{R12}$ correction requires the reduced density matrix from the FCIQMC calculation and therefore suffers from the initiator error as well.
However, the error in the $[2]_{R12}$ correction is small compared to the basis set correlation energy and decays quickly with respect to the total number of walkers.
However, it also has the opposite sign from the error in the FCIQMC energy, and thus slightly reduces the overall initiator error at low walker
numbers due to favourable error cancellation.

\section{Conclusions}

The canonical transcorrelation approach of Yanai and Shiozaki and the $[2]_{R12}$ approach of Valeev {\it et al.} constitute two contrasting philosophies in the
attempt to create a universal, explicitly correlated approach to basis set incompleteness in quantum chemistry. In the former, the dynamic correlation from the cusps
is included first, to produce an effective, two-body Hamiltonian operator, to which the multireference strong correlation methods can be applied. This has the advantage
that the multireference wavefunction capturing the static correlation in the system can relax in the presence of the geminal amplitudes and dynamic correlation
contributions. Furthermore, we showed that the transcorrelation reduces the computational effort of the FCIQMC calculation as it improves the convergence of the
initiator error with respect to the total number of walkers. The CT method is also
in principle a state-universal approach, in that the specifics of the high-energy geminal contributions to the effective Hamiltonian should be relatively insensitive to the
state or specifics of the choice of geminal. However, one source of ambiguity in the construction of $\hat{H}_{\text{TC}}$ comes from the choice of the active
space that is required for the CASSCF calculation of the initial 'trial' density, and for the transcorrelation itself.  


Contrasting this, the $[2]_{R12}$ perturbatively couples the geminals directly to the multireference wavefunction, which provides geminal relaxation in a 
state-specific fashion and without any initial trial wavefunction, but conversely does not allow for relaxation of the CI amplitudes in the presence of dynamic correlation
component of the electronic structure. However, these effects are likely to be minor details in the largely single-reference systems studied here, and it can be seen
that in both methods, the FCIQMC-R12 and CT-FCIQMC methods both achieve accurate results even in the small basis sets, with the CT approach slightly outperforming the $[2]_{R12}$ method.
Computational requirements for the integration of the $[2]_{R12}$ methodology with FCIQMC are larger, due to the
necessity of sampling the two-body density matrix. The sampling
at least doubles the memory and processing cost of the FCIQMC calculation due to the requirement of two replicas.

However, aside from the cost of the density matrix sampling, the computational effort of both methods is very cheap compared to the costs of the FCIQMC
calculations, scaling as $\mathcal{O}[N^6]$ with the system size. The different calculations for the molecules of the G1
standard set took between a few minutes and about two days on a single core for both of the explicitly correlated treatments. 
Conclusively,  with little additional computational effort, explicitly correlated FCIQMC
calculations in small basis sets can achieve the same accuracy that is normally only achieved with
considerably more expensive FCIQMC calculations in larger basis sets. 
This broadens the applicability of the FCIQMC method considerably and allows accurate calculations of larger systems, where in the future, we will consider problems with stronger correlation,
and applications to excited states and solid state systems\cite{Booth_PlaneW,PhysRevLett.115.066402,Usvyat_2013}.

\section{Acknowledgements}
We would like to sincerely thank Takeshi Yanai for kindly allowing us to use his code \texttt{ORZ}, and are also grateful to Luke Roskop for his help with the
\texttt{MPQC}-\texttt{GAMESS} interface.
G.H.B gratefully acknowledges funding from the Royal Society.
The work has been funded by the EPSRC under Grant No.: EP/J003867/1.

%


\begin{thebibliography}{126}%
\makeatletter
\providecommand \@ifxundefined [1]{%
 \@ifx{#1\undefined}
}%
\providecommand \@ifnum [1]{%
 \ifnum #1\expandafter \@firstoftwo
 \else \expandafter \@secondoftwo
 \fi
}%
\providecommand \@ifx [1]{%
 \ifx #1\expandafter \@firstoftwo
 \else \expandafter \@secondoftwo
 \fi
}%
\providecommand \natexlab [1]{#1}%
\providecommand \enquote  [1]{``#1''}%
\providecommand \bibnamefont  [1]{#1}%
\providecommand \bibfnamefont [1]{#1}%
\providecommand \citenamefont [1]{#1}%
\providecommand \href@noop [0]{\@secondoftwo}%
\providecommand \href [0]{\begingroup \@sanitize@url \@href}%
\providecommand \@href[1]{\@@startlink{#1}\@@href}%
\providecommand \@@href[1]{\endgroup#1\@@endlink}%
\providecommand \@sanitize@url [0]{\catcode `\\12\catcode `\$12\catcode
  `\&12\catcode `\#12\catcode `\^12\catcode `\_12\catcode `\%12\relax}%
\providecommand \@@startlink[1]{}%
\providecommand \@@endlink[0]{}%
\providecommand \url  [0]{\begingroup\@sanitize@url \@url }%
\providecommand \@url [1]{\endgroup\@href {#1}{\urlprefix }}%
\providecommand \urlprefix  [0]{URL }%
\providecommand \Eprint [0]{\href }%
\providecommand \doibase [0]{http://dx.doi.org/}%
\providecommand \selectlanguage [0]{\@gobble}%
\providecommand \bibinfo  [0]{\@secondoftwo}%
\providecommand \bibfield  [0]{\@secondoftwo}%
\providecommand \translation [1]{[#1]}%
\providecommand \BibitemOpen [0]{}%
\providecommand \bibitemStop [0]{}%
\providecommand \bibitemNoStop [0]{.\EOS\space}%
\providecommand \EOS [0]{\spacefactor3000\relax}%
\providecommand \BibitemShut  [1]{\csname bibitem#1\endcsname}%
\let\auto@bib@innerbib\@empty
\bibitem [{\citenamefont {Booth}, \citenamefont {Thom},\ and\ \citenamefont
  {Alavi}(2009)}]{Booth_Alavi_2009}%
  \BibitemOpen
  \bibfield  {author} {\bibinfo {author} {\bibfnamefont {G.~H.}\ \bibnamefont
  {Booth}}, \bibinfo {author} {\bibfnamefont {A.~J.~W.}\ \bibnamefont {Thom}},
  \ and\ \bibinfo {author} {\bibfnamefont {A.}~\bibnamefont {Alavi}},\
  }\href@noop {} {\bibfield  {journal} {\bibinfo  {journal} {J. Chem. Phys.}\
  }\textbf {\bibinfo {volume} {131}},\ \bibinfo {pages} {054106} (\bibinfo
  {year} {2009})}\BibitemShut {NoStop}%
\bibitem [{\citenamefont {Booth}\ and\ \citenamefont
  {Alavi}(2010)}]{Booth_Alavi_2010}%
  \BibitemOpen
  \bibfield  {author} {\bibinfo {author} {\bibfnamefont {G.~H.}\ \bibnamefont
  {Booth}}\ and\ \bibinfo {author} {\bibfnamefont {A.}~\bibnamefont {Alavi}},\
  }\href@noop {} {\bibfield  {journal} {\bibinfo  {journal} {J. Chem. Phys.}\
  }\textbf {\bibinfo {volume} {132}},\ \bibinfo {pages} {174104} (\bibinfo
  {year} {2010})}\BibitemShut {NoStop}%
\bibitem [{\citenamefont {Booth}\ \emph {et~al.}(2011)\citenamefont {Booth},
  \citenamefont {Cleland}, \citenamefont {Thom},\ and\ \citenamefont
  {Alavi}}]{Booth_Alavi_2011}%
  \BibitemOpen
  \bibfield  {author} {\bibinfo {author} {\bibfnamefont {G.~H.}\ \bibnamefont
  {Booth}}, \bibinfo {author} {\bibfnamefont {D.}~\bibnamefont {Cleland}},
  \bibinfo {author} {\bibfnamefont {A.~J.~W.}\ \bibnamefont {Thom}}, \ and\
  \bibinfo {author} {\bibfnamefont {A.}~\bibnamefont {Alavi}},\ }\href@noop {}
  {\bibfield  {journal} {\bibinfo  {journal} {J. Chem. Phys.}\ }\textbf
  {\bibinfo {volume} {135}},\ \bibinfo {pages} {084104} (\bibinfo {year}
  {2011})}\BibitemShut {NoStop}%
\bibitem [{\citenamefont {Thomas}\ \emph {et~al.}(2014)\citenamefont {Thomas},
  \citenamefont {Overy}, \citenamefont {Booth},\ and\ \citenamefont
  {Alavi}}]{FCIQMC_N2}%
  \BibitemOpen
  \bibfield  {author} {\bibinfo {author} {\bibfnamefont {R.~E.}\ \bibnamefont
  {Thomas}}, \bibinfo {author} {\bibfnamefont {C.}~\bibnamefont {Overy}},
  \bibinfo {author} {\bibfnamefont {G.~H.}\ \bibnamefont {Booth}}, \ and\
  \bibinfo {author} {\bibfnamefont {A.}~\bibnamefont {Alavi}},\ }\href@noop {}
  {\bibfield  {journal} {\bibinfo  {journal} {J. Chem. Theor. Comput.}\
  }\textbf {\bibinfo {volume} {10}},\ \bibinfo {pages} {1915} (\bibinfo {year}
  {2014})}\BibitemShut {NoStop}%
\bibitem [{\citenamefont {Daday}\ \emph {et~al.}(2012)\citenamefont {Daday},
  \citenamefont {Smart}, \citenamefont {Booth}, \citenamefont {Alavi},\ and\
  \citenamefont {Filippi}}]{FCIQMC_ETHENE}%
  \BibitemOpen
  \bibfield  {author} {\bibinfo {author} {\bibfnamefont {C.}~\bibnamefont
  {Daday}}, \bibinfo {author} {\bibfnamefont {S.}~\bibnamefont {Smart}},
  \bibinfo {author} {\bibfnamefont {G.~H.}\ \bibnamefont {Booth}}, \bibinfo
  {author} {\bibfnamefont {A.}~\bibnamefont {Alavi}}, \ and\ \bibinfo {author}
  {\bibfnamefont {C.}~\bibnamefont {Filippi}},\ }\href@noop {} {\bibfield
  {journal} {\bibinfo  {journal} {J. Chem. Theor. Comput.}\ }\textbf {\bibinfo
  {volume} {8}},\ \bibinfo {pages} {4441} (\bibinfo {year} {2012})}\BibitemShut
  {NoStop}%
\bibitem [{\citenamefont {Cleland}\ \emph {et~al.}(2012)\citenamefont
  {Cleland}, \citenamefont {Booth}, \citenamefont {Overy},\ and\ \citenamefont
  {Alavi}}]{FCIQMC_diatomics}%
  \BibitemOpen
  \bibfield  {author} {\bibinfo {author} {\bibfnamefont {D.}~\bibnamefont
  {Cleland}}, \bibinfo {author} {\bibfnamefont {G.~H.}\ \bibnamefont {Booth}},
  \bibinfo {author} {\bibfnamefont {C.}~\bibnamefont {Overy}}, \ and\ \bibinfo
  {author} {\bibfnamefont {A.}~\bibnamefont {Alavi}},\ }\href@noop {}
  {\bibfield  {journal} {\bibinfo  {journal} {J. Chem. Theor. Comput.}\
  }\textbf {\bibinfo {volume} {8}},\ \bibinfo {pages} {4138} (\bibinfo {year}
  {2012})}\BibitemShut {NoStop}%
\bibitem [{\citenamefont {Thomas}, \citenamefont {Booth},\ and\ \citenamefont
  {Alavi}(2015)}]{Thomas_2015}%
  \BibitemOpen
  \bibfield  {author} {\bibinfo {author} {\bibfnamefont {R.~E.}\ \bibnamefont
  {Thomas}}, \bibinfo {author} {\bibfnamefont {G.~H.}\ \bibnamefont {Booth}}, \
  and\ \bibinfo {author} {\bibfnamefont {A.}~\bibnamefont {Alavi}},\
  }\href@noop {} {\bibfield  {journal} {\bibinfo  {journal} {Phys. Rev. Lett.}\
  }\textbf {\bibinfo {volume} {114}},\ \bibinfo {pages} {033001} (\bibinfo
  {year} {2015})}\BibitemShut {NoStop}%
\bibitem [{\citenamefont {Schwarz}, \citenamefont {Booth},\ and\ \citenamefont
  {Alavi}(2015)}]{FCIQMC_Hubbard}%
  \BibitemOpen
  \bibfield  {author} {\bibinfo {author} {\bibfnamefont {L.~R.}\ \bibnamefont
  {Schwarz}}, \bibinfo {author} {\bibfnamefont {G.~H.}\ \bibnamefont {Booth}},
  \ and\ \bibinfo {author} {\bibfnamefont {A.}~\bibnamefont {Alavi}},\ }\href
  {\doibase 10.1103/PhysRevB.91.045139} {\bibfield  {journal} {\bibinfo
  {journal} {Phys. Rev. B}\ }\textbf {\bibinfo {volume} {91}},\ \bibinfo
  {pages} {045139} (\bibinfo {year} {2015})}\BibitemShut {NoStop}%
\bibitem [{\citenamefont {Shepherd}\ \emph
  {et~al.}(2012{\natexlab{a}})\citenamefont {Shepherd}, \citenamefont {Booth},
  \citenamefont {Gr\"uneis},\ and\ \citenamefont
  {Alavi}}]{Shepherd_Alavi_2012_1}%
  \BibitemOpen
  \bibfield  {author} {\bibinfo {author} {\bibfnamefont {J.~J.}\ \bibnamefont
  {Shepherd}}, \bibinfo {author} {\bibfnamefont {G.~H.}\ \bibnamefont {Booth}},
  \bibinfo {author} {\bibfnamefont {A.}~\bibnamefont {Gr\"uneis}}, \ and\
  \bibinfo {author} {\bibfnamefont {A.}~\bibnamefont {Alavi}},\ }\href@noop {}
  {\bibfield  {journal} {\bibinfo  {journal} {Phys. Rev. B}\ }\textbf {\bibinfo
  {volume} {85}},\ \bibinfo {pages} {081103(R)} (\bibinfo {year}
  {2012}{\natexlab{a}})}\BibitemShut {NoStop}%
\bibitem [{\citenamefont {Booth}\ \emph {et~al.}(2013)\citenamefont {Booth},
  \citenamefont {Gruneis}, \citenamefont {Kresse},\ and\ \citenamefont
  {Alavi}}]{FCIQMC_solidstates}%
  \BibitemOpen
  \bibfield  {author} {\bibinfo {author} {\bibfnamefont {G.~H.}\ \bibnamefont
  {Booth}}, \bibinfo {author} {\bibfnamefont {A.}~\bibnamefont {Gruneis}},
  \bibinfo {author} {\bibfnamefont {G.}~\bibnamefont {Kresse}}, \ and\ \bibinfo
  {author} {\bibfnamefont {A.}~\bibnamefont {Alavi}},\ }\href
  {http://dx.doi.org/10.1038/nature11770} {\bibfield  {journal} {\bibinfo
  {journal} {Nature}\ }\textbf {\bibinfo {volume} {493}},\ \bibinfo {pages}
  {365} (\bibinfo {year} {2013})}\BibitemShut {NoStop}%
\bibitem [{\citenamefont {Cleland}, \citenamefont {Booth},\ and\ \citenamefont
  {Alavi}(2010)}]{Cleland_Alavi_2010}%
  \BibitemOpen
  \bibfield  {author} {\bibinfo {author} {\bibfnamefont {D.}~\bibnamefont
  {Cleland}}, \bibinfo {author} {\bibfnamefont {G.~H.}\ \bibnamefont {Booth}},
  \ and\ \bibinfo {author} {\bibfnamefont {A.}~\bibnamefont {Alavi}},\ }\href
  {\doibase http://dx.doi.org/10.1063/1.3302277} {\bibfield  {journal}
  {\bibinfo  {journal} {J. Chem. Phys.}\ }\textbf {\bibinfo {volume} {132}},\
  \bibinfo {eid} {041103} (\bibinfo {year} {2010})}\BibitemShut {NoStop}%
\bibitem [{\citenamefont {Cleland}, \citenamefont {Booth},\ and\ \citenamefont
  {Alavi}(2011)}]{Cleland_Alavi_2011}%
  \BibitemOpen
  \bibfield  {author} {\bibinfo {author} {\bibfnamefont {D.~M.}\ \bibnamefont
  {Cleland}}, \bibinfo {author} {\bibfnamefont {G.~H.}\ \bibnamefont {Booth}},
  \ and\ \bibinfo {author} {\bibfnamefont {A.}~\bibnamefont {Alavi}},\ }\href
  {\doibase http://dx.doi.org/10.1063/1.3525712} {\bibfield  {journal}
  {\bibinfo  {journal} {J. Chem. Phys.}\ }\textbf {\bibinfo {volume} {134}},\
  \bibinfo {eid} {024112} (\bibinfo {year} {2011})}\BibitemShut {NoStop}%
\bibitem [{\citenamefont {Petruzielo}\ \emph {et~al.}(2012)\citenamefont
  {Petruzielo}, \citenamefont {Holmes}, \citenamefont {Changlani},
  \citenamefont {Nightingale},\ and\ \citenamefont
  {Umrigar}}]{semi_stochastic}%
  \BibitemOpen
  \bibfield  {author} {\bibinfo {author} {\bibfnamefont {F.~R.}\ \bibnamefont
  {Petruzielo}}, \bibinfo {author} {\bibfnamefont {A.~A.}\ \bibnamefont
  {Holmes}}, \bibinfo {author} {\bibfnamefont {H.~J.}\ \bibnamefont
  {Changlani}}, \bibinfo {author} {\bibfnamefont {M.~P.}\ \bibnamefont
  {Nightingale}}, \ and\ \bibinfo {author} {\bibfnamefont {C.~J.}\ \bibnamefont
  {Umrigar}},\ }\href {\doibase 10.1103/PhysRevLett.109.230201} {\bibfield
  {journal} {\bibinfo  {journal} {Phys. Rev. Lett.}\ }\textbf {\bibinfo
  {volume} {109}},\ \bibinfo {pages} {230201} (\bibinfo {year}
  {2012})}\BibitemShut {NoStop}%
\bibitem [{\citenamefont {Blunt}\ \emph
  {et~al.}(2015{\natexlab{a}})\citenamefont {Blunt}, \citenamefont {Smart},
  \citenamefont {Kersten}, \citenamefont {Spencer}, \citenamefont {Booth},\
  and\ \citenamefont {Alavi}}]{Blunt_SS}%
  \BibitemOpen
  \bibfield  {author} {\bibinfo {author} {\bibfnamefont {N.~S.}\ \bibnamefont
  {Blunt}}, \bibinfo {author} {\bibfnamefont {S.~D.}\ \bibnamefont {Smart}},
  \bibinfo {author} {\bibfnamefont {J.~A.~F.}\ \bibnamefont {Kersten}},
  \bibinfo {author} {\bibfnamefont {J.~S.}\ \bibnamefont {Spencer}}, \bibinfo
  {author} {\bibfnamefont {G.~H.}\ \bibnamefont {Booth}}, \ and\ \bibinfo
  {author} {\bibfnamefont {A.}~\bibnamefont {Alavi}},\ }\href {\doibase
  http://dx.doi.org/10.1063/1.4920975} {\bibfield  {journal} {\bibinfo
  {journal} {J. Chem. Phys.}\ }\textbf {\bibinfo {volume} {142}},\ \bibinfo
  {eid} {184107} (\bibinfo {year} {2015}{\natexlab{a}})}\BibitemShut {NoStop}%
\bibitem [{\citenamefont {Humeniuk}\ and\ \citenamefont
  {Mitri\'{c}}(2014)}]{excited_states_QMC}%
  \BibitemOpen
  \bibfield  {author} {\bibinfo {author} {\bibfnamefont {A.}~\bibnamefont
  {Humeniuk}}\ and\ \bibinfo {author} {\bibfnamefont {R.}~\bibnamefont
  {Mitri\'{c}}},\ }\href {\doibase http://dx.doi.org/10.1063/1.4901020}
  {\bibfield  {journal} {\bibinfo  {journal} {J. Chem. Phys.}\ }\textbf
  {\bibinfo {volume} {141}},\ \bibinfo {pages} {194104} (\bibinfo {year}
  {2014})}\BibitemShut {NoStop}%
\bibitem [{\citenamefont {Blunt}, \citenamefont {Alavi},\ and\ \citenamefont
  {Booth}(2015)}]{Blunt_2015}%
  \BibitemOpen
  \bibfield  {author} {\bibinfo {author} {\bibfnamefont {N.~S.}\ \bibnamefont
  {Blunt}}, \bibinfo {author} {\bibfnamefont {A.}~\bibnamefont {Alavi}}, \ and\
  \bibinfo {author} {\bibfnamefont {G.~H.}\ \bibnamefont {Booth}},\ }\href@noop
  {} {\bibfield  {journal} {\bibinfo  {journal} {Phys. Rev. Lett.}\ }\textbf
  {\bibinfo {volume} {115}},\ \bibinfo {pages} {050603} (\bibinfo {year}
  {2015})}\BibitemShut {NoStop}%
\bibitem [{\citenamefont {Booth}\ and\ \citenamefont {Chan}(2012)}]{Chan_2012}%
  \BibitemOpen
  \bibfield  {author} {\bibinfo {author} {\bibfnamefont {G.~H.}\ \bibnamefont
  {Booth}}\ and\ \bibinfo {author} {\bibfnamefont {G.-L.}\ \bibnamefont
  {Chan}},\ }\href@noop {} {\bibfield  {journal} {\bibinfo  {journal} {J. Chem.
  Phys.}\ }\textbf {\bibinfo {volume} {137}},\ \bibinfo {pages} {191102}
  (\bibinfo {year} {2012})}\BibitemShut {NoStop}%
\bibitem [{\citenamefont {Ohtsuka}\ and\ \citenamefont
  {Nagase}(2011)}]{Nagase2011}%
  \BibitemOpen
  \bibfield  {author} {\bibinfo {author} {\bibfnamefont {Y.}~\bibnamefont
  {Ohtsuka}}\ and\ \bibinfo {author} {\bibfnamefont {S.}~\bibnamefont
  {Nagase}},\ }\href@noop {} {\bibfield  {journal} {\bibinfo  {journal} {Theor.
  Chem. Acc.}\ }\textbf {\bibinfo {volume} {130}},\ \bibinfo {pages} {501}
  (\bibinfo {year} {2011})}\BibitemShut {NoStop}%
\bibitem [{\citenamefont {Ten-no}(2013)}]{Ten-no_MSQMC}%
  \BibitemOpen
  \bibfield  {author} {\bibinfo {author} {\bibfnamefont {S.}~\bibnamefont
  {Ten-no}},\ }\href@noop {} {\bibfield  {journal} {\bibinfo  {journal} {J.
  Chem. Phys.}\ }\textbf {\bibinfo {volume} {138}},\ \bibinfo {eid} {164126}
  (\bibinfo {year} {2013})}\BibitemShut {NoStop}%
\bibitem [{\citenamefont {Blunt}\ \emph
  {et~al.}(2015{\natexlab{b}})\citenamefont {Blunt}, \citenamefont {Smart},
  \citenamefont {Booth},\ and\ \citenamefont {Alavi}}]{blunt2015excited}%
  \BibitemOpen
  \bibfield  {author} {\bibinfo {author} {\bibfnamefont {N.}~\bibnamefont
  {Blunt}}, \bibinfo {author} {\bibfnamefont {S.~D.}\ \bibnamefont {Smart}},
  \bibinfo {author} {\bibfnamefont {G.~H.}\ \bibnamefont {Booth}}, \ and\
  \bibinfo {author} {\bibfnamefont {A.}~\bibnamefont {Alavi}},\ }\href@noop {}
  {\bibfield  {journal} {\bibinfo  {journal} {J. Chem. Phys.}\ }\textbf
  {\bibinfo {volume} {143}},\ \bibinfo {pages} {134117} (\bibinfo {year}
  {2015}{\natexlab{b}})}\BibitemShut {NoStop}%
\bibitem [{\citenamefont {Blunt}\ \emph {et~al.}(2014)\citenamefont {Blunt},
  \citenamefont {Rogers}, \citenamefont {Spencer},\ and\ \citenamefont
  {Foulkes}}]{DensityMatrix_Blunt}%
  \BibitemOpen
  \bibfield  {author} {\bibinfo {author} {\bibfnamefont {N.~S.}\ \bibnamefont
  {Blunt}}, \bibinfo {author} {\bibfnamefont {T.~W.}\ \bibnamefont {Rogers}},
  \bibinfo {author} {\bibfnamefont {J.~S.}\ \bibnamefont {Spencer}}, \ and\
  \bibinfo {author} {\bibfnamefont {W.~M.~C.}\ \bibnamefont {Foulkes}},\ }\href
  {\doibase 10.1103/PhysRevB.89.245124} {\bibfield  {journal} {\bibinfo
  {journal} {Phys. Rev. B}\ }\textbf {\bibinfo {volume} {89}},\ \bibinfo
  {pages} {245124} (\bibinfo {year} {2014})}\BibitemShut {NoStop}%
\bibitem [{\citenamefont {Overy}\ \emph {et~al.}(2014)\citenamefont {Overy},
  \citenamefont {Booth}, \citenamefont {Blunt}, \citenamefont {Shepherd},
  \citenamefont {Cleland},\ and\ \citenamefont {Alavi}}]{RDM_Overy}%
  \BibitemOpen
  \bibfield  {author} {\bibinfo {author} {\bibfnamefont {C.}~\bibnamefont
  {Overy}}, \bibinfo {author} {\bibfnamefont {G.~H.}\ \bibnamefont {Booth}},
  \bibinfo {author} {\bibfnamefont {N.~S.}\ \bibnamefont {Blunt}}, \bibinfo
  {author} {\bibfnamefont {J.~J.}\ \bibnamefont {Shepherd}}, \bibinfo {author}
  {\bibfnamefont {D.}~\bibnamefont {Cleland}}, \ and\ \bibinfo {author}
  {\bibfnamefont {A.}~\bibnamefont {Alavi}},\ }\href@noop {} {\bibfield
  {journal} {\bibinfo  {journal} {J. Chem. Phys.}\ }\textbf {\bibinfo {volume}
  {141}},\ \bibinfo {eid} {244117} (\bibinfo {year} {2014})}\BibitemShut
  {NoStop}%
\bibitem [{\citenamefont {Thomas}\ \emph {et~al.}(2015)\citenamefont {Thomas},
  \citenamefont {Sun}, \citenamefont {Alavi},\ and\ \citenamefont
  {Booth}}]{Thomas_CASSCF}%
  \BibitemOpen
  \bibfield  {author} {\bibinfo {author} {\bibfnamefont {R.~E.}\ \bibnamefont
  {Thomas}}, \bibinfo {author} {\bibfnamefont {Q.}~\bibnamefont {Sun}},
  \bibinfo {author} {\bibfnamefont {A.}~\bibnamefont {Alavi}}, \ and\ \bibinfo
  {author} {\bibfnamefont {G.~H.}\ \bibnamefont {Booth}},\ }\href@noop {}
  {\bibfield  {journal} {\bibinfo  {journal} {J. Chem. Theor. Comput.}\
  }\textbf {\bibinfo {volume} {11}},\ \bibinfo {pages} {5316} (\bibinfo {year}
  {2015})}\BibitemShut {NoStop}%
\bibitem [{\citenamefont {Manni}, \citenamefont {Smart},\ and\ \citenamefont
  {Alavi}(2016)}]{Manni_2016}%
  \BibitemOpen
  \bibfield  {author} {\bibinfo {author} {\bibfnamefont {G.~L.}\ \bibnamefont
  {Manni}}, \bibinfo {author} {\bibfnamefont {S.~D.}\ \bibnamefont {Smart}}, \
  and\ \bibinfo {author} {\bibfnamefont {A.}~\bibnamefont {Alavi}},\
  }\href@noop {} {\bibfield  {journal} {\bibinfo  {journal} {J. Chem. Theor.
  Comput.}\ }\textbf {\bibinfo {volume} {12}},\ \bibinfo {pages} {1245}
  (\bibinfo {year} {2016})}\BibitemShut {NoStop}%
\bibitem [{\citenamefont {Schwartz}(1962)}]{L3_convergence1}%
  \BibitemOpen
  \bibfield  {author} {\bibinfo {author} {\bibfnamefont {C.}~\bibnamefont
  {Schwartz}},\ }\href {\doibase 10.1103/PhysRev.126.1015} {\bibfield
  {journal} {\bibinfo  {journal} {Phys. Rev.}\ }\textbf {\bibinfo {volume}
  {126}},\ \bibinfo {pages} {1015} (\bibinfo {year} {1962})}\BibitemShut
  {NoStop}%
\bibitem [{\citenamefont {Hill}(1985)}]{L3_convergence2}%
  \BibitemOpen
  \bibfield  {author} {\bibinfo {author} {\bibfnamefont {R.~N.}\ \bibnamefont
  {Hill}},\ }\href {\doibase http://dx.doi.org/10.1063/1.449481} {\bibfield
  {journal} {\bibinfo  {journal} {J. Chem. Phys.}\ }\textbf {\bibinfo {volume}
  {83}},\ \bibinfo {pages} {1173} (\bibinfo {year} {1985})}\BibitemShut
  {NoStop}%
\bibitem [{\citenamefont {Kutzelnigg}\ and\ \citenamefont
  {Morgan~III}(1992)}]{Morgan_1992}%
  \BibitemOpen
  \bibfield  {author} {\bibinfo {author} {\bibfnamefont {W.}~\bibnamefont
  {Kutzelnigg}}\ and\ \bibinfo {author} {\bibfnamefont {J.}~\bibnamefont
  {Morgan~III}},\ }\href@noop {} {\bibfield  {journal} {\bibinfo  {journal} {J.
  Chem. Phys.}\ }\textbf {\bibinfo {volume} {96}},\ \bibinfo {pages} {4484}
  (\bibinfo {year} {1992})}\BibitemShut {NoStop}%
\bibitem [{\citenamefont {Shepherd}\ \emph
  {et~al.}(2012{\natexlab{b}})\citenamefont {Shepherd}, \citenamefont
  {Gr\"uneis}, \citenamefont {Booth}, \citenamefont {Kresse},\ and\
  \citenamefont {Alavi}}]{Shepherd_Alavi_2012_2}%
  \BibitemOpen
  \bibfield  {author} {\bibinfo {author} {\bibfnamefont {J.~J.}\ \bibnamefont
  {Shepherd}}, \bibinfo {author} {\bibfnamefont {A.}~\bibnamefont {Gr\"uneis}},
  \bibinfo {author} {\bibfnamefont {G.~H.}\ \bibnamefont {Booth}}, \bibinfo
  {author} {\bibfnamefont {G.}~\bibnamefont {Kresse}}, \ and\ \bibinfo {author}
  {\bibfnamefont {A.}~\bibnamefont {Alavi}},\ }\href@noop {} {\bibfield
  {journal} {\bibinfo  {journal} {Phys. Rev. B}\ }\textbf {\bibinfo {volume}
  {86}},\ \bibinfo {pages} {035111} (\bibinfo {year}
  {2012}{\natexlab{b}})}\BibitemShut {NoStop}%
\bibitem [{\citenamefont {Slater}(1928)}]{Slater.31.333}%
  \BibitemOpen
  \bibfield  {author} {\bibinfo {author} {\bibfnamefont {J.~C.}\ \bibnamefont
  {Slater}},\ }\href {\doibase 10.1103/PhysRev.31.333} {\bibfield  {journal}
  {\bibinfo  {journal} {Phys. Rev.}\ }\textbf {\bibinfo {volume} {31}},\
  \bibinfo {pages} {333} (\bibinfo {year} {1928})}\BibitemShut {NoStop}%
\bibitem [{\citenamefont {Hylleraas}(1929)}]{Hylleraas_1929}%
  \BibitemOpen
  \bibfield  {author} {\bibinfo {author} {\bibfnamefont {E.~A.}\ \bibnamefont
  {Hylleraas}},\ }\href@noop {} {\bibfield  {journal} {\bibinfo  {journal}
  {Zeitschrift f{\"u}r Physik}\ }\textbf {\bibinfo {volume} {54}},\ \bibinfo
  {pages} {347} (\bibinfo {year} {1929})}\BibitemShut {NoStop}%
\bibitem [{\citenamefont {Pack}\ and\ \citenamefont
  {Brown}(1966)}]{Pack_cusp1}%
  \BibitemOpen
  \bibfield  {author} {\bibinfo {author} {\bibfnamefont {R.~T.}\ \bibnamefont
  {Pack}}\ and\ \bibinfo {author} {\bibfnamefont {W.~B.}\ \bibnamefont
  {Brown}},\ }\href {\doibase http://dx.doi.org/10.1063/1.1727605} {\bibfield
  {journal} {\bibinfo  {journal} {J. Chem. Phys.}\ }\textbf {\bibinfo {volume}
  {45}},\ \bibinfo {pages} {556} (\bibinfo {year} {1966})}\BibitemShut
  {NoStop}%
\bibitem [{\citenamefont {Kato}(1957)}]{Kato_1957}%
  \BibitemOpen
  \bibfield  {author} {\bibinfo {author} {\bibfnamefont {T.}~\bibnamefont
  {Kato}},\ }\href {\doibase 10.1002/cpa.3160100201} {\bibfield  {journal}
  {\bibinfo  {journal} {Communications on Pure and Applied Mathematics}\
  }\textbf {\bibinfo {volume} {10}},\ \bibinfo {pages} {151} (\bibinfo {year}
  {1957})}\BibitemShut {NoStop}%
\bibitem [{\citenamefont {Klopper}\ \emph {et~al.}(2006)\citenamefont
  {Klopper}, \citenamefont {Manby}, \citenamefont {Ten-No},\ and\ \citenamefont
  {Valeev}}]{R12_rev1}%
  \BibitemOpen
  \bibfield  {author} {\bibinfo {author} {\bibfnamefont {W.}~\bibnamefont
  {Klopper}}, \bibinfo {author} {\bibfnamefont {F.~R.}\ \bibnamefont {Manby}},
  \bibinfo {author} {\bibfnamefont {S.}~\bibnamefont {Ten-No}}, \ and\ \bibinfo
  {author} {\bibfnamefont {E.~F.}\ \bibnamefont {Valeev}},\ }\href@noop {}
  {\bibfield  {journal} {\bibinfo  {journal} {Int. Rev. Phys. Chem.}\ }\textbf
  {\bibinfo {volume} {25}},\ \bibinfo {pages} {427} (\bibinfo {year}
  {2006})}\BibitemShut {NoStop}%
\bibitem [{\citenamefont {Kong}, \citenamefont {Bischoff},\ and\ \citenamefont
  {Valeev}(2012)}]{Kong_Valeev_2012}%
  \BibitemOpen
  \bibfield  {author} {\bibinfo {author} {\bibfnamefont {L.}~\bibnamefont
  {Kong}}, \bibinfo {author} {\bibfnamefont {F.~A.}\ \bibnamefont {Bischoff}},
  \ and\ \bibinfo {author} {\bibfnamefont {E.~F.}\ \bibnamefont {Valeev}},\
  }\href@noop {} {\bibfield  {journal} {\bibinfo  {journal} {Chem. Rev.}\
  }\textbf {\bibinfo {volume} {112}},\ \bibinfo {pages} {75 } (\bibinfo {year}
  {2012})}\BibitemShut {NoStop}%
\bibitem [{\citenamefont {H\"attig}\ \emph {et~al.}(2012)\citenamefont
  {H\"attig}, \citenamefont {Klopper}, \citenamefont {K\"ohn},\ and\
  \citenamefont {Tew}}]{rev_Tew}%
  \BibitemOpen
  \bibfield  {author} {\bibinfo {author} {\bibfnamefont {C.}~\bibnamefont
  {H\"attig}}, \bibinfo {author} {\bibfnamefont {W.}~\bibnamefont {Klopper}},
  \bibinfo {author} {\bibfnamefont {A.}~\bibnamefont {K\"ohn}}, \ and\ \bibinfo
  {author} {\bibfnamefont {D.~P.}\ \bibnamefont {Tew}},\ }\href@noop {}
  {\bibfield  {journal} {\bibinfo  {journal} {Chem. Rev.}\ }\textbf {\bibinfo
  {volume} {112}},\ \bibinfo {pages} {4} (\bibinfo {year} {2012})}\BibitemShut
  {NoStop}%
\bibitem [{\citenamefont {Kutzelnigg}(1985)}]{R12_Kutzelnigg}%
  \BibitemOpen
  \bibfield  {author} {\bibinfo {author} {\bibfnamefont {W.}~\bibnamefont
  {Kutzelnigg}},\ }\href {\doibase 10.1007/BF00527669} {\bibfield  {journal}
  {\bibinfo  {journal} {Theoretica chimica acta}\ }\textbf {\bibinfo {volume}
  {68}},\ \bibinfo {pages} {445} (\bibinfo {year} {1985})}\BibitemShut
  {NoStop}%
\bibitem [{\citenamefont {Termath}, \citenamefont {Klopper},\ and\
  \citenamefont {Kutzelnigg}(1991)}]{Kutzelnigg_MP2r12}%
  \BibitemOpen
  \bibfield  {author} {\bibinfo {author} {\bibfnamefont {V.}~\bibnamefont
  {Termath}}, \bibinfo {author} {\bibfnamefont {W.}~\bibnamefont {Klopper}}, \
  and\ \bibinfo {author} {\bibfnamefont {W.}~\bibnamefont {Kutzelnigg}},\
  }\href {\doibase http://dx.doi.org/10.1063/1.459922} {\bibfield  {journal}
  {\bibinfo  {journal} {J. Chem. Phys.}\ }\textbf {\bibinfo {volume} {94}},\
  \bibinfo {pages} {2002} (\bibinfo {year} {1991})}\BibitemShut {NoStop}%
\bibitem [{\citenamefont {Kutzelnigg}\ and\ \citenamefont
  {Klopper}(1991)}]{Kutzelnigg_standardA}%
  \BibitemOpen
  \bibfield  {author} {\bibinfo {author} {\bibfnamefont {W.}~\bibnamefont
  {Kutzelnigg}}\ and\ \bibinfo {author} {\bibfnamefont {W.}~\bibnamefont
  {Klopper}},\ }\href {\doibase http://dx.doi.org/10.1063/1.459921} {\bibfield
  {journal} {\bibinfo  {journal} {J. Chem. Phys.}\ }\textbf {\bibinfo {volume}
  {94}},\ \bibinfo {pages} {1985} (\bibinfo {year} {1991})}\BibitemShut
  {NoStop}%
\bibitem [{\citenamefont {Noga}, \citenamefont {Kutzelnigg},\ and\
  \citenamefont {Klopper}(1992)}]{Noga_CCR12}%
  \BibitemOpen
  \bibfield  {author} {\bibinfo {author} {\bibfnamefont {J.}~\bibnamefont
  {Noga}}, \bibinfo {author} {\bibfnamefont {W.}~\bibnamefont {Kutzelnigg}}, \
  and\ \bibinfo {author} {\bibfnamefont {W.}~\bibnamefont {Klopper}},\ }\href
  {\doibase http://dx.doi.org/10.1016/0009-2614(92)87034-M} {\bibfield
  {journal} {\bibinfo  {journal} {Chem. Phys. Lett.}\ }\textbf {\bibinfo
  {volume} {199}},\ \bibinfo {pages} {497 } (\bibinfo {year}
  {1992})}\BibitemShut {NoStop}%
\bibitem [{\citenamefont {Noga}\ and\ \citenamefont
  {Kutzelnigg}(1994)}]{Noga_CCR12_2}%
  \BibitemOpen
  \bibfield  {author} {\bibinfo {author} {\bibfnamefont {J.}~\bibnamefont
  {Noga}}\ and\ \bibinfo {author} {\bibfnamefont {W.}~\bibnamefont
  {Kutzelnigg}},\ }\href {\doibase http://dx.doi.org/10.1063/1.468266}
  {\bibfield  {journal} {\bibinfo  {journal} {J. Chem. Phys.}\ }\textbf
  {\bibinfo {volume} {101}},\ \bibinfo {pages} {7738} (\bibinfo {year}
  {1994})}\BibitemShut {NoStop}%
\bibitem [{\citenamefont {Klopper}\ and\ \citenamefont {Samson}(2002)}]{ABS}%
  \BibitemOpen
  \bibfield  {author} {\bibinfo {author} {\bibfnamefont {W.}~\bibnamefont
  {Klopper}}\ and\ \bibinfo {author} {\bibfnamefont {C.~C.~M.}\ \bibnamefont
  {Samson}},\ }\href {\doibase http://dx.doi.org/10.1063/1.1461814} {\bibfield
  {journal} {\bibinfo  {journal} {J. Chem. Phys.}\ }\textbf {\bibinfo {volume}
  {116}},\ \bibinfo {pages} {6397} (\bibinfo {year} {2002})}\BibitemShut
  {NoStop}%
\bibitem [{\citenamefont {Valeev}(2004)}]{Valeev2004190}%
  \BibitemOpen
  \bibfield  {author} {\bibinfo {author} {\bibfnamefont {E.~F.}\ \bibnamefont
  {Valeev}},\ }\href {\doibase http://dx.doi.org/10.1016/j.cplett.2004.07.061}
  {\bibfield  {journal} {\bibinfo  {journal} {Chem. Phys. Lett.}\ }\textbf
  {\bibinfo {volume} {395}},\ \bibinfo {pages} {190 } (\bibinfo {year}
  {2004})}\BibitemShut {NoStop}%
\bibitem [{\citenamefont {May}\ and\ \citenamefont {Manby}(2004)}]{DF1}%
  \BibitemOpen
  \bibfield  {author} {\bibinfo {author} {\bibfnamefont {A.~J.}\ \bibnamefont
  {May}}\ and\ \bibinfo {author} {\bibfnamefont {F.~R.}\ \bibnamefont
  {Manby}},\ }\href {\doibase 10.1063/1.1780891} {\bibfield  {journal}
  {\bibinfo  {journal} {J. Chem. Phys.}\ }\textbf {\bibinfo {volume} {121}},\
  \bibinfo {pages} {4479} (\bibinfo {year} {2004})}\BibitemShut {NoStop}%
\bibitem [{\citenamefont {Manby}(2003)}]{DF2}%
  \BibitemOpen
  \bibfield  {author} {\bibinfo {author} {\bibfnamefont {F.~R.}\ \bibnamefont
  {Manby}},\ }\href {\doibase http://dx.doi.org/10.1063/1.1594713} {\bibfield
  {journal} {\bibinfo  {journal} {J. Chem. Phys.}\ }\textbf {\bibinfo {volume}
  {119}},\ \bibinfo {pages} {4607} (\bibinfo {year} {2003})}\BibitemShut
  {NoStop}%
\bibitem [{\citenamefont {Ten-no}(2007{\natexlab{a}})}]{Tenno_2007}%
  \BibitemOpen
  \bibfield  {author} {\bibinfo {author} {\bibfnamefont {S.}~\bibnamefont
  {Ten-no}},\ }\href {\doibase http://dx.doi.org/10.1063/1.2403853} {\bibfield
  {journal} {\bibinfo  {journal} {J. Chem. Phys.}\ }\textbf {\bibinfo {volume}
  {126}},\ \bibinfo {eid} {014108} (\bibinfo {year}
  {2007}{\natexlab{a}})}\BibitemShut {NoStop}%
\bibitem [{\citenamefont {Ten-no}(2000{\natexlab{a}})}]{Tenno_2000_1}%
  \BibitemOpen
  \bibfield  {author} {\bibinfo {author} {\bibfnamefont {S.}~\bibnamefont
  {Ten-no}},\ }\href {\doibase 10.1016/S0009-2614(00)01066-6} {\bibfield
  {journal} {\bibinfo  {journal} {Chem. Phys. Lett.}\ }\textbf {\bibinfo
  {volume} {330}},\ \bibinfo {pages} {169 } (\bibinfo {year}
  {2000}{\natexlab{a}})}\BibitemShut {NoStop}%
\bibitem [{\citenamefont {Ten-no}(2000{\natexlab{b}})}]{Tenno_2000_2}%
  \BibitemOpen
  \bibfield  {author} {\bibinfo {author} {\bibfnamefont {S.}~\bibnamefont
  {Ten-no}},\ }\href {\doibase 10.1016/S0009-2614(00)01067-8} {\bibfield
  {journal} {\bibinfo  {journal} {Chem. Phys. Lett.}\ }\textbf {\bibinfo
  {volume} {330}},\ \bibinfo {pages} {175 } (\bibinfo {year}
  {2000}{\natexlab{b}})}\BibitemShut {NoStop}%
\bibitem [{\citenamefont {Ten-no}(2004)}]{Tenno_2004}%
  \BibitemOpen
  \bibfield  {author} {\bibinfo {author} {\bibfnamefont {S.}~\bibnamefont
  {Ten-no}},\ }\href {\doibase 10.1016/j.cplett.2004.09.041} {\bibfield
  {journal} {\bibinfo  {journal} {Chem. Phys. Lett.}\ }\textbf {\bibinfo
  {volume} {398}},\ \bibinfo {pages} {56} (\bibinfo {year} {2004})}\BibitemShut
  {NoStop}%
\bibitem [{\citenamefont {Tew}\ and\ \citenamefont
  {Klopper}(2005)}]{Tew_geminal}%
  \BibitemOpen
  \bibfield  {author} {\bibinfo {author} {\bibfnamefont {D.~P.}\ \bibnamefont
  {Tew}}\ and\ \bibinfo {author} {\bibfnamefont {W.}~\bibnamefont {Klopper}},\
  }\href {\doibase 10.1063/1.1999632} {\bibfield  {journal} {\bibinfo
  {journal} {J. Chem. Phys.}\ }\textbf {\bibinfo {volume} {123}},\ \bibinfo
  {pages} {074101} (\bibinfo {year} {2005})}\BibitemShut {NoStop}%
\bibitem [{\citenamefont {May}\ \emph {et~al.}(2005)\citenamefont {May},
  \citenamefont {Valeev}, \citenamefont {Polly},\ and\ \citenamefont
  {Manby}}]{r12f12}%
  \BibitemOpen
  \bibfield  {author} {\bibinfo {author} {\bibfnamefont {A.~J.}\ \bibnamefont
  {May}}, \bibinfo {author} {\bibfnamefont {E.}~\bibnamefont {Valeev}},
  \bibinfo {author} {\bibfnamefont {R.}~\bibnamefont {Polly}}, \ and\ \bibinfo
  {author} {\bibfnamefont {F.~R.}\ \bibnamefont {Manby}},\ }\href {\doibase
  10.1039/B507781H} {\bibfield  {journal} {\bibinfo  {journal} {Phys. Chem.
  Chem. Phys.}\ }\textbf {\bibinfo {volume} {7}},\ \bibinfo {pages} {2710}
  (\bibinfo {year} {2005})}\BibitemShut {NoStop}%
\bibitem [{\citenamefont {Ten-no}(2007{\natexlab{b}})}]{Tenno_MRF12}%
  \BibitemOpen
  \bibfield  {author} {\bibinfo {author} {\bibfnamefont {S.}~\bibnamefont
  {Ten-no}},\ }\href {\doibase http://dx.doi.org/10.1016/j.cplett.2007.09.006}
  {\bibfield  {journal} {\bibinfo  {journal} {Chem. Phys. Lett.}\ }\textbf
  {\bibinfo {volume} {447}},\ \bibinfo {pages} {175 } (\bibinfo {year}
  {2007}{\natexlab{b}})}\BibitemShut {NoStop}%
\bibitem [{\citenamefont {Gdanitz}(1999)}]{Gdanitz_MRCI-R12}%
  \BibitemOpen
  \bibfield  {author} {\bibinfo {author} {\bibfnamefont {R.~J.}\ \bibnamefont
  {Gdanitz}},\ }\href {\doibase
  http://dx.doi.org/10.1016/S0009-2614(99)00985-9} {\bibfield  {journal}
  {\bibinfo  {journal} {Chem. Phys. Lett.}\ }\textbf {\bibinfo {volume}
  {312}},\ \bibinfo {pages} {578 } (\bibinfo {year} {1999})}\BibitemShut
  {NoStop}%
\bibitem [{\citenamefont {Shiozaki}, \citenamefont {Knizia},\ and\
  \citenamefont {Werner}(2011)}]{Shiozaki_MRCIF12}%
  \BibitemOpen
  \bibfield  {author} {\bibinfo {author} {\bibfnamefont {T.}~\bibnamefont
  {Shiozaki}}, \bibinfo {author} {\bibfnamefont {G.}~\bibnamefont {Knizia}}, \
  and\ \bibinfo {author} {\bibfnamefont {H.-J.}\ \bibnamefont {Werner}},\
  }\href {\doibase http://dx.doi.org/10.1063/1.3528720} {\bibfield  {journal}
  {\bibinfo  {journal} {J. Chem. Phys.}\ }\textbf {\bibinfo {volume} {134}},\
  \bibinfo {eid} {034113} (\bibinfo {year} {2011})}\BibitemShut {NoStop}%
\bibitem [{\citenamefont {Shiozaki}\ and\ \citenamefont
  {Werner}(2010)}]{Werner_MRF12}%
  \BibitemOpen
  \bibfield  {author} {\bibinfo {author} {\bibfnamefont {T.}~\bibnamefont
  {Shiozaki}}\ and\ \bibinfo {author} {\bibfnamefont {H.-J.}\ \bibnamefont
  {Werner}},\ }\href {\doibase http://dx.doi.org/10.1063/1.3489000} {\bibfield
  {journal} {\bibinfo  {journal} {J. Chem. Phys.}\ }\textbf {\bibinfo {volume}
  {133}},\ \bibinfo {pages} {141103} (\bibinfo {year} {2010})}\BibitemShut
  {NoStop}%
\bibitem [{\citenamefont {Torheyden}\ and\ \citenamefont
  {Valeev}(2009)}]{Torheyden_R12}%
  \BibitemOpen
  \bibfield  {author} {\bibinfo {author} {\bibfnamefont {M.}~\bibnamefont
  {Torheyden}}\ and\ \bibinfo {author} {\bibfnamefont {E.~F.}\ \bibnamefont
  {Valeev}},\ }\href {\doibase http://dx.doi.org/10.1063/1.3254836} {\bibfield
  {journal} {\bibinfo  {journal} {J. Chem. Phys.}\ }\textbf {\bibinfo {volume}
  {131}},\ \bibinfo {eid} {171103} (\bibinfo {year} {2009})}\BibitemShut
  {NoStop}%
\bibitem [{\citenamefont {Kong}\ and\ \citenamefont
  {Valeev}(2011)}]{Kong_Valeev_2011}%
  \BibitemOpen
  \bibfield  {author} {\bibinfo {author} {\bibfnamefont {L.}~\bibnamefont
  {Kong}}\ and\ \bibinfo {author} {\bibfnamefont {E.~F.}\ \bibnamefont
  {Valeev}},\ }\href@noop {} {\bibfield  {journal} {\bibinfo  {journal} {J.
  Chem. Phys.}\ }\textbf {\bibinfo {volume} {135}},\ \bibinfo {pages} {214105}
  (\bibinfo {year} {2011})}\BibitemShut {NoStop}%
\bibitem [{\citenamefont {Yanai}\ and\ \citenamefont
  {Shiozaki}(2012)}]{Yanai_Shiozaki_2012}%
  \BibitemOpen
  \bibfield  {author} {\bibinfo {author} {\bibfnamefont {T.}~\bibnamefont
  {Yanai}}\ and\ \bibinfo {author} {\bibfnamefont {T.}~\bibnamefont
  {Shiozaki}},\ }\href@noop {} {\bibfield  {journal} {\bibinfo  {journal} {J.
  Chem. Phys.}\ }\textbf {\bibinfo {volume} {136}} (\bibinfo {year}
  {2012})}\BibitemShut {NoStop}%
\bibitem [{\citenamefont {Yanai}\ and\ \citenamefont {Chan}(2007)}]{CT_Yanai}%
  \BibitemOpen
  \bibfield  {author} {\bibinfo {author} {\bibfnamefont {T.}~\bibnamefont
  {Yanai}}\ and\ \bibinfo {author} {\bibfnamefont {G.~K.-L.}\ \bibnamefont
  {Chan}},\ }\href {\doibase http://dx.doi.org/10.1063/1.2761870} {\bibfield
  {journal} {\bibinfo  {journal} {J. Chem. Phys.}\ }\textbf {\bibinfo {volume}
  {127}},\ \bibinfo {eid} {104107} (\bibinfo {year} {2007})}\BibitemShut
  {NoStop}%
\bibitem [{\citenamefont {Booth}\ \emph {et~al.}(2012)\citenamefont {Booth},
  \citenamefont {Cleland}, \citenamefont {Alavi},\ and\ \citenamefont
  {Tew}}]{Booth_Tew_2012}%
  \BibitemOpen
  \bibfield  {author} {\bibinfo {author} {\bibfnamefont {G.~H.}\ \bibnamefont
  {Booth}}, \bibinfo {author} {\bibfnamefont {D.}~\bibnamefont {Cleland}},
  \bibinfo {author} {\bibfnamefont {A.}~\bibnamefont {Alavi}}, \ and\ \bibinfo
  {author} {\bibfnamefont {D.~P.}\ \bibnamefont {Tew}},\ }\href {\doibase
  http://dx.doi.org/10.1063/1.4762445} {\bibfield  {journal} {\bibinfo
  {journal} {J. Chem. Phys.}\ }\textbf {\bibinfo {volume} {137}},\ \bibinfo
  {eid} {164112} (\bibinfo {year} {2012})}\BibitemShut {NoStop}%
\bibitem [{\citenamefont {Sharma}\ \emph {et~al.}(2014)\citenamefont {Sharma},
  \citenamefont {Yanai}, \citenamefont {Booth}, \citenamefont {Umrigar},\ and\
  \citenamefont {Chan}}]{FCIQMC_CT}%
  \BibitemOpen
  \bibfield  {author} {\bibinfo {author} {\bibfnamefont {S.}~\bibnamefont
  {Sharma}}, \bibinfo {author} {\bibfnamefont {T.}~\bibnamefont {Yanai}},
  \bibinfo {author} {\bibfnamefont {G.~H.}\ \bibnamefont {Booth}}, \bibinfo
  {author} {\bibfnamefont {C.~J.}\ \bibnamefont {Umrigar}}, \ and\ \bibinfo
  {author} {\bibfnamefont {G.~K.-L.}\ \bibnamefont {Chan}},\ }\href {\doibase
  http://dx.doi.org/10.1063/1.4867383} {\bibfield  {journal} {\bibinfo
  {journal} {J. Chem. Phys.}\ }\textbf {\bibinfo {volume} {140}},\ \bibinfo
  {eid} {104112} (\bibinfo {year} {2014})}\BibitemShut {NoStop}%
\bibitem [{\citenamefont {Pople}\ \emph {et~al.}(1989)\citenamefont {Pople},
  \citenamefont {Head-Gordon}, \citenamefont {Fox}, \citenamefont
  {Raghavachari},\ and\ \citenamefont {Curtiss}}]{G1_set_1}%
  \BibitemOpen
  \bibfield  {author} {\bibinfo {author} {\bibfnamefont {J.~A.}\ \bibnamefont
  {Pople}}, \bibinfo {author} {\bibfnamefont {M.}~\bibnamefont {Head-Gordon}},
  \bibinfo {author} {\bibfnamefont {D.~J.}\ \bibnamefont {Fox}}, \bibinfo
  {author} {\bibfnamefont {K.}~\bibnamefont {Raghavachari}}, \ and\ \bibinfo
  {author} {\bibfnamefont {L.~A.}\ \bibnamefont {Curtiss}},\ }\href {\doibase
  http://dx.doi.org/10.1063/1.456415} {\bibfield  {journal} {\bibinfo
  {journal} {J. Chem. Phys.}\ }\textbf {\bibinfo {volume} {90}},\ \bibinfo
  {pages} {5622} (\bibinfo {year} {1989})}\BibitemShut {NoStop}%
\bibitem [{\citenamefont {Curtiss}\ \emph {et~al.}(1990)\citenamefont
  {Curtiss}, \citenamefont {Jones}, \citenamefont {Trucks}, \citenamefont
  {Raghavachari},\ and\ \citenamefont {Pople}}]{G1_set_2}%
  \BibitemOpen
  \bibfield  {author} {\bibinfo {author} {\bibfnamefont {L.~A.}\ \bibnamefont
  {Curtiss}}, \bibinfo {author} {\bibfnamefont {C.}~\bibnamefont {Jones}},
  \bibinfo {author} {\bibfnamefont {G.~W.}\ \bibnamefont {Trucks}}, \bibinfo
  {author} {\bibfnamefont {K.}~\bibnamefont {Raghavachari}}, \ and\ \bibinfo
  {author} {\bibfnamefont {J.~A.}\ \bibnamefont {Pople}},\ }\href {\doibase
  http://dx.doi.org/10.1063/1.458892} {\bibfield  {journal} {\bibinfo
  {journal} {J. Chem. Phys.}\ }\textbf {\bibinfo {volume} {93}},\ \bibinfo
  {pages} {2537} (\bibinfo {year} {1990})}\BibitemShut {NoStop}%
\bibitem [{\citenamefont {Booth}, \citenamefont {Smart},\ and\ \citenamefont
  {Alavi}(2014)}]{Booth_algorithm}%
  \BibitemOpen
  \bibfield  {author} {\bibinfo {author} {\bibfnamefont {G.~H.}\ \bibnamefont
  {Booth}}, \bibinfo {author} {\bibfnamefont {S.~D.}\ \bibnamefont {Smart}}, \
  and\ \bibinfo {author} {\bibfnamefont {A.}~\bibnamefont {Alavi}},\ }\href
  {\doibase 10.1080/00268976.2013.877165} {\bibfield  {journal} {\bibinfo
  {journal} {Molecular Physics}\ }\textbf {\bibinfo {volume} {112}},\ \bibinfo
  {pages} {1855} (\bibinfo {year} {2014})}\BibitemShut {NoStop}%
\bibitem [{\citenamefont {Roskop}\ \emph {et~al.}(2014)\citenamefont {Roskop},
  \citenamefont {Kong}, \citenamefont {Valeev}, \citenamefont {Gordon},\ and\
  \citenamefont {Windus}}]{Luke_MPQC}%
  \BibitemOpen
  \bibfield  {author} {\bibinfo {author} {\bibfnamefont {L.~B.}\ \bibnamefont
  {Roskop}}, \bibinfo {author} {\bibfnamefont {L.}~\bibnamefont {Kong}},
  \bibinfo {author} {\bibfnamefont {E.~F.}\ \bibnamefont {Valeev}}, \bibinfo
  {author} {\bibfnamefont {M.~S.}\ \bibnamefont {Gordon}}, \ and\ \bibinfo
  {author} {\bibfnamefont {T.~L.}\ \bibnamefont {Windus}},\ }\href@noop {}
  {\bibfield  {journal} {\bibinfo  {journal} {J. Chem. Theor. Comput.}\
  }\textbf {\bibinfo {volume} {10}},\ \bibinfo {pages} {90} (\bibinfo {year}
  {2014})}\BibitemShut {NoStop}%
\bibitem [{\citenamefont {Kutzelnigg}\ and\ \citenamefont
  {Mukherjee}(1997)}]{Mukherjee}%
  \BibitemOpen
  \bibfield  {author} {\bibinfo {author} {\bibfnamefont {W.}~\bibnamefont
  {Kutzelnigg}}\ and\ \bibinfo {author} {\bibfnamefont {D.}~\bibnamefont
  {Mukherjee}},\ }\href {\doibase http://dx.doi.org/10.1063/1.474405}
  {\bibfield  {journal} {\bibinfo  {journal} {J. Chem. Phys.}\ }\textbf
  {\bibinfo {volume} {107}},\ \bibinfo {pages} {432} (\bibinfo {year}
  {1997})}\BibitemShut {NoStop}%
\bibitem [{\citenamefont {Kutzelnigg}, \citenamefont {Shamasundar},\ and\
  \citenamefont {Mukherjee}(2010)}]{Kutzelnigg_NO}%
  \BibitemOpen
  \bibfield  {author} {\bibinfo {author} {\bibfnamefont {W.}~\bibnamefont
  {Kutzelnigg}}, \bibinfo {author} {\bibfnamefont {K.~R.}\ \bibnamefont
  {Shamasundar}}, \ and\ \bibinfo {author} {\bibfnamefont {D.}~\bibnamefont
  {Mukherjee}},\ }\href@noop {} {\bibfield  {journal} {\bibinfo  {journal}
  {Mol. Phys.}\ }\textbf {\bibinfo {volume} {108}},\ \bibinfo {pages} {433}
  (\bibinfo {year} {2010})}\BibitemShut {NoStop}%
\bibitem [{\citenamefont {Kutzelnigg}(2006)}]{DCFT_Kutzelnigg}%
  \BibitemOpen
  \bibfield  {author} {\bibinfo {author} {\bibfnamefont {W.}~\bibnamefont
  {Kutzelnigg}},\ }\href@noop {} {\bibfield  {journal} {\bibinfo  {journal} {J.
  Chem. Phys.}\ }\textbf {\bibinfo {volume} {125}},\ \bibinfo {pages} {171101}
  (\bibinfo {year} {2006})}\BibitemShut {NoStop}%
\bibitem [{Note1()}]{Note1}%
  \BibitemOpen
  \bibinfo {note} {The FCIQMC code can be obtained from {\protect \tt
  https://github.com/ghb24/NECI\protect \_STABLE.git}}\BibitemShut {NoStop}%
\bibitem [{\citenamefont {Janssen}\ \emph {et~al.}(2006)\citenamefont
  {Janssen}, \citenamefont {Nielsen}, \citenamefont {Leininger}, \citenamefont
  {Valeev}, \citenamefont {Kenny},\ and\ \citenamefont {Seidl}}]{mpqc1}%
  \BibitemOpen
  \bibfield  {author} {\bibinfo {author} {\bibfnamefont {C.~L.}\ \bibnamefont
  {Janssen}}, \bibinfo {author} {\bibfnamefont {I.~B.}\ \bibnamefont
  {Nielsen}}, \bibinfo {author} {\bibfnamefont {M.~L.}\ \bibnamefont
  {Leininger}}, \bibinfo {author} {\bibfnamefont {E.~F.}\ \bibnamefont
  {Valeev}}, \bibinfo {author} {\bibfnamefont {J.~P.}\ \bibnamefont {Kenny}}, \
  and\ \bibinfo {author} {\bibfnamefont {E.~T.}\ \bibnamefont {Seidl}},\ }\href
  {http://www.mpqc.org} {\enquote {\bibinfo {title} {The massively parallel
  quantum chemistry program ({MPQC}), version 2.4.0 prerelease},}\ } (\bibinfo
  {year} {2006})\BibitemShut {NoStop}%
\bibitem [{\citenamefont {Janssen}, \citenamefont {Seidl},\ and\ \citenamefont
  {Colvin}(1995)}]{mpqc2}%
  \BibitemOpen
  \bibfield  {author} {\bibinfo {author} {\bibfnamefont {C.}~\bibnamefont
  {Janssen}}, \bibinfo {author} {\bibfnamefont {E.}~\bibnamefont {Seidl}}, \
  and\ \bibinfo {author} {\bibfnamefont {M.}~\bibnamefont {Colvin}},\ }in\
  \href@noop {} {\emph {\bibinfo {booktitle} {ACS Symposium Series, Parallel
  Computing in Computational Chemistry}}},\ Vol.\ \bibinfo {volume} {592}\
  (\bibinfo  {publisher} {American Chemical Society},\ \bibinfo {year}
  {1995})\BibitemShut {NoStop}%
\bibitem [{\citenamefont {Hirschfelder}(1963)}]{hirschfelder1963}%
  \BibitemOpen
  \bibfield  {author} {\bibinfo {author} {\bibfnamefont {J.~O.}\ \bibnamefont
  {Hirschfelder}},\ }\href@noop {} {\bibfield  {journal} {\bibinfo  {journal}
  {J. Chem. Phys.}\ }\textbf {\bibinfo {volume} {39}},\ \bibinfo {pages} {3145}
  (\bibinfo {year} {1963})}\BibitemShut {NoStop}%
\bibitem [{\citenamefont {Boys}\ and\ \citenamefont
  {Handy}(1969{\natexlab{a}})}]{Boys_Handy_1969_1}%
  \BibitemOpen
  \bibfield  {author} {\bibinfo {author} {\bibfnamefont {S.~F.}\ \bibnamefont
  {Boys}}\ and\ \bibinfo {author} {\bibfnamefont {N.~C.}\ \bibnamefont
  {Handy}},\ }\href@noop {} {\bibfield  {journal} {\bibinfo  {journal} {Proc.
  R. Soc. Lond. A}\ }\textbf {\bibinfo {volume} {309}},\ \bibinfo {pages} {209}
  (\bibinfo {year} {1969}{\natexlab{a}})}\BibitemShut {NoStop}%
\bibitem [{\citenamefont {Boys}\ and\ \citenamefont
  {Handy}(1969{\natexlab{b}})}]{Boys_Handy_1969_2}%
  \BibitemOpen
  \bibfield  {author} {\bibinfo {author} {\bibfnamefont {S.~F.}\ \bibnamefont
  {Boys}}\ and\ \bibinfo {author} {\bibfnamefont {N.~C.}\ \bibnamefont
  {Handy}},\ }\href {\doibase doi: 10.1098/rspa.1969.0061} {\bibfield
  {journal} {\bibinfo  {journal} {Proc. R. Soc. Lond. A}\ }\textbf {\bibinfo
  {volume} {310}},\ \bibinfo {pages} {43} (\bibinfo {year}
  {1969}{\natexlab{b}})}\BibitemShut {NoStop}%
\bibitem [{\citenamefont {Boys}\ and\ \citenamefont
  {Handy}(1969{\natexlab{c}})}]{Boys_Handy_1969_3}%
  \BibitemOpen
  \bibfield  {author} {\bibinfo {author} {\bibfnamefont {S.~F.}\ \bibnamefont
  {Boys}}\ and\ \bibinfo {author} {\bibfnamefont {N.~C.}\ \bibnamefont
  {Handy}},\ }\href {\doibase 10.1098/rspa.1969.0062} {\bibfield  {journal}
  {\bibinfo  {journal} {Proc. R. Soc. Lond. A}\ }\textbf {\bibinfo {volume}
  {310}},\ \bibinfo {pages} {63} (\bibinfo {year}
  {1969}{\natexlab{c}})}\BibitemShut {NoStop}%
\bibitem [{\citenamefont {Umezawa}\ and\ \citenamefont
  {Tsuneyuki}(2003)}]{Umezawa_Tsuneyuki_2003}%
  \BibitemOpen
  \bibfield  {author} {\bibinfo {author} {\bibfnamefont {N.}~\bibnamefont
  {Umezawa}}\ and\ \bibinfo {author} {\bibfnamefont {S.}~\bibnamefont
  {Tsuneyuki}},\ }\href@noop {} {\bibfield  {journal} {\bibinfo  {journal} {J.
  of Chem. Phys.}\ }\textbf {\bibinfo {volume} {119}},\ \bibinfo {pages}
  {10015} (\bibinfo {year} {2003})}\BibitemShut {NoStop}%
\bibitem [{\citenamefont {Umezawa}\ and\ \citenamefont
  {Tsuneyuki}(2004)}]{Umezawa_Tsuneyuki_2004}%
  \BibitemOpen
  \bibfield  {author} {\bibinfo {author} {\bibfnamefont {N.}~\bibnamefont
  {Umezawa}}\ and\ \bibinfo {author} {\bibfnamefont {S.}~\bibnamefont
  {Tsuneyuki}},\ }\href {\doibase 10.1103/PhysRevB.69.165.102} {\bibfield
  {journal} {\bibinfo  {journal} {Phys. Rev. B}\ }\textbf {\bibinfo {volume}
  {69}} (\bibinfo {year} {2004}),\ 10.1103/PhysRevB.69.165.102}\BibitemShut
  {NoStop}%
\bibitem [{\citenamefont {Luo}(2010)}]{Luo_2010}%
  \BibitemOpen
  \bibfield  {author} {\bibinfo {author} {\bibfnamefont {H.}~\bibnamefont
  {Luo}},\ }\href@noop {} {\bibfield  {journal} {\bibinfo  {journal} {J. Chem.
  Phys.}\ }\textbf {\bibinfo {volume} {133}},\ \bibinfo {pages} {154109}
  (\bibinfo {year} {2010})}\BibitemShut {NoStop}%
\bibitem [{\citenamefont {Luo}(2011)}]{Luo_2011}%
  \BibitemOpen
  \bibfield  {author} {\bibinfo {author} {\bibfnamefont {H.}~\bibnamefont
  {Luo}},\ }\href@noop {} {\bibfield  {journal} {\bibinfo  {journal} {J. Chem.
  Phys.}\ }\textbf {\bibinfo {volume} {135}},\ \bibinfo {pages} {024109}
  (\bibinfo {year} {2011})}\BibitemShut {NoStop}%
\bibitem [{\citenamefont {Luo}(2012)}]{Luo_2012}%
  \BibitemOpen
  \bibfield  {author} {\bibinfo {author} {\bibfnamefont {H.}~\bibnamefont
  {Luo}},\ }\href@noop {} {\bibfield  {journal} {\bibinfo  {journal} {J. Chem.
  Phys.}\ }\textbf {\bibinfo {volume} {136}},\ \bibinfo {pages} {224111}
  (\bibinfo {year} {2012})}\BibitemShut {NoStop}%
\bibitem [{\citenamefont {White}(2002)}]{White_2002}%
  \BibitemOpen
  \bibfield  {author} {\bibinfo {author} {\bibfnamefont {S.~R.}\ \bibnamefont
  {White}},\ }\href@noop {} {\bibfield  {journal} {\bibinfo  {journal} {J.
  Chem. Phys.}\ }\textbf {\bibinfo {volume} {117}},\ \bibinfo {pages} {7472}
  (\bibinfo {year} {2002})}\BibitemShut {NoStop}%
\bibitem [{\citenamefont {Neuscamman}, \citenamefont {Yanai},\ and\
  \citenamefont {Chan}(2010{\natexlab{a}})}]{Neuscamman_2010}%
  \BibitemOpen
  \bibfield  {author} {\bibinfo {author} {\bibfnamefont {E.}~\bibnamefont
  {Neuscamman}}, \bibinfo {author} {\bibfnamefont {T.}~\bibnamefont {Yanai}}, \
  and\ \bibinfo {author} {\bibfnamefont {G.~K.-L.}\ \bibnamefont {Chan}},\
  }\href@noop {} {\bibfield  {journal} {\bibinfo  {journal} {Int. Rev. Phys.
  Chem.}\ }\textbf {\bibinfo {volume} {29}},\ \bibinfo {pages} {231} (\bibinfo
  {year} {2010}{\natexlab{a}})}\BibitemShut {NoStop}%
\bibitem [{\citenamefont {Yanai}\ and\ \citenamefont
  {Chan}(2006)}]{Yanai_2006}%
  \BibitemOpen
  \bibfield  {author} {\bibinfo {author} {\bibfnamefont {T.}~\bibnamefont
  {Yanai}}\ and\ \bibinfo {author} {\bibfnamefont {G.~K.-L.}\ \bibnamefont
  {Chan}},\ }\href@noop {} {\bibfield  {journal} {\bibinfo  {journal} {J. Chem.
  Phys.}\ }\textbf {\bibinfo {volume} {124}},\ \bibinfo {pages} {194106}
  (\bibinfo {year} {2006})}\BibitemShut {NoStop}%
\bibitem [{\citenamefont {Chan}\ and\ \citenamefont
  {Yanai}(2007)}]{Yanai_2007_2}%
  \BibitemOpen
  \bibfield  {author} {\bibinfo {author} {\bibfnamefont {G.~K.-L.}\
  \bibnamefont {Chan}}\ and\ \bibinfo {author} {\bibfnamefont {T.}~\bibnamefont
  {Yanai}},\ }\href@noop {} {\bibfield  {journal} {\bibinfo  {journal} {Adv.
  Chem. Phys.}\ }\textbf {\bibinfo {volume} {134}},\ \bibinfo {pages} {343}
  (\bibinfo {year} {2007})}\BibitemShut {NoStop}%
\bibitem [{\citenamefont {Neuscamman}, \citenamefont {Yanai},\ and\
  \citenamefont {Chan}(2009)}]{Neuscamman_09}%
  \BibitemOpen
  \bibfield  {author} {\bibinfo {author} {\bibfnamefont {E.}~\bibnamefont
  {Neuscamman}}, \bibinfo {author} {\bibfnamefont {T.}~\bibnamefont {Yanai}}, \
  and\ \bibinfo {author} {\bibfnamefont {G.~K.-L.}\ \bibnamefont {Chan}},\
  }\href@noop {} {\bibfield  {journal} {\bibinfo  {journal} {J. Chem. Phys.}\
  }\textbf {\bibinfo {volume} {130}},\ \bibinfo {pages} {124102} (\bibinfo
  {year} {2009})}\BibitemShut {NoStop}%
\bibitem [{\citenamefont {Neuscamman}, \citenamefont {Yanai},\ and\
  \citenamefont {Chan}(2010{\natexlab{b}})}]{Neuscamman_10}%
  \BibitemOpen
  \bibfield  {author} {\bibinfo {author} {\bibfnamefont {E.}~\bibnamefont
  {Neuscamman}}, \bibinfo {author} {\bibfnamefont {T.}~\bibnamefont {Yanai}}, \
  and\ \bibinfo {author} {\bibfnamefont {G.~K.-L.}\ \bibnamefont {Chan}},\
  }\href@noop {} {\bibfield  {journal} {\bibinfo  {journal} {J. Chem. Phys.}\
  }\textbf {\bibinfo {volume} {132}},\ \bibinfo {pages} {024106} (\bibinfo
  {year} {2010}{\natexlab{b}})}\BibitemShut {NoStop}%
\bibitem [{\citenamefont {Yanai}\ \emph {et~al.}(2010)\citenamefont {Yanai},
  \citenamefont {Kurashige}, \citenamefont {Neuscamman},\ and\ \citenamefont
  {Chan}}]{Neuscamman_10_2}%
  \BibitemOpen
  \bibfield  {author} {\bibinfo {author} {\bibfnamefont {T.}~\bibnamefont
  {Yanai}}, \bibinfo {author} {\bibfnamefont {Y.}~\bibnamefont {Kurashige}},
  \bibinfo {author} {\bibfnamefont {E.}~\bibnamefont {Neuscamman}}, \ and\
  \bibinfo {author} {\bibfnamefont {G.~K.-L.}\ \bibnamefont {Chan}},\
  }\href@noop {} {\bibfield  {journal} {\bibinfo  {journal} {J. Chem. Phys.}\
  }\textbf {\bibinfo {volume} {132}},\ \bibinfo {pages} {024105} (\bibinfo
  {year} {2010})}\BibitemShut {NoStop}%
\bibitem [{\citenamefont {Yanai}\ \emph {et~al.}(2012)\citenamefont {Yanai},
  \citenamefont {Kurashige}, \citenamefont {Neuscamman},\ and\ \citenamefont
  {Chan}}]{Yanai_12}%
  \BibitemOpen
  \bibfield  {author} {\bibinfo {author} {\bibfnamefont {T.}~\bibnamefont
  {Yanai}}, \bibinfo {author} {\bibfnamefont {Y.}~\bibnamefont {Kurashige}},
  \bibinfo {author} {\bibfnamefont {E.}~\bibnamefont {Neuscamman}}, \ and\
  \bibinfo {author} {\bibfnamefont {G.~K.-L.}\ \bibnamefont {Chan}},\
  }\href@noop {} {\bibfield  {journal} {\bibinfo  {journal} {Phys. Chem. Chem.
  Phys.}\ }\textbf {\bibinfo {volume} {14}},\ \bibinfo {pages} {7809} (\bibinfo
  {year} {2012})}\BibitemShut {NoStop}%
\bibitem [{\citenamefont {Watson}\ and\ \citenamefont
  {Chan}(2016)}]{Watson_16}%
  \BibitemOpen
  \bibfield  {author} {\bibinfo {author} {\bibfnamefont {T.}~\bibnamefont
  {Watson}}\ and\ \bibinfo {author} {\bibfnamefont {G.~K.-L.}\ \bibnamefont
  {Chan}},\ }\href@noop {} {\bibfield  {journal} {\bibinfo  {journal} {J. Chem.
  Theor. Comput.}\ }\textbf {\bibinfo {volume} {12}},\ \bibinfo {pages} {512}
  (\bibinfo {year} {2016})}\BibitemShut {NoStop}%
\bibitem [{\citenamefont {Ked\v{z}uch}, \citenamefont {Milko},\ and\
  \citenamefont {Noga}(2005)}]{SA_C}%
  \BibitemOpen
  \bibfield  {author} {\bibinfo {author} {\bibfnamefont {S.}~\bibnamefont
  {Ked\v{z}uch}}, \bibinfo {author} {\bibfnamefont {M.}~\bibnamefont {Milko}},
  \ and\ \bibinfo {author} {\bibfnamefont {J.}~\bibnamefont {Noga}},\ }\href
  {\doibase 10.1002/qua.20744} {\bibfield  {journal} {\bibinfo  {journal} {Int.
  J. Quant. Chem.}\ }\textbf {\bibinfo {volume} {105}},\ \bibinfo {pages} {929}
  (\bibinfo {year} {2005})}\BibitemShut {NoStop}%
\bibitem [{\citenamefont {Shiozaki}(2009)}]{Shiozaki2009CPL}%
  \BibitemOpen
  \bibfield  {author} {\bibinfo {author} {\bibfnamefont {T.}~\bibnamefont
  {Shiozaki}},\ }\href@noop {} {\bibfield  {journal} {\bibinfo  {journal}
  {{Chem. Phys. Lett.}}\ }\textbf {\bibinfo {volume} {479}},\ \bibinfo {pages}
  {160} (\bibinfo {year} {2009})}\BibitemShut {NoStop}%
\bibitem [{\citenamefont {Shiozaki}(2011)}]{Shiozaki2011JUQ}%
  \BibitemOpen
  \bibfield  {author} {\bibinfo {author} {\bibfnamefont {T.}~\bibnamefont
  {Shiozaki}},\ }\href@noop {} {\bibfield  {journal} {\bibinfo  {journal} {{J.
  Unsolved Questions}}\ }\textbf {\bibinfo {volume} {1}},\ \bibinfo {pages} {1
  (http://junq.info/?p=348, accessed on Dec. 30, 2011)} (\bibinfo {year}
  {2011})}\BibitemShut {NoStop}%
\bibitem [{\citenamefont {Shiozaki}(2014)}]{Shiozaki2009program}%
  \BibitemOpen
  \bibfield  {author} {\bibinfo {author} {\bibfnamefont {T.}~\bibnamefont
  {Shiozaki}},\ }\href@noop {} {} (\bibinfo {year} {2014}),\ \bibinfo {note}
  {{{\sc PolyR12}, {\it the program for explicitly correlated
  electron-correlation calculations of molecules and solids}, can be obtained
  from https://github.com/nubakery/libslater}}\BibitemShut {NoStop}%
\bibitem [{\citenamefont {Feller}, \citenamefont {Peterson},\ and\
  \citenamefont {Dixon}(2008)}]{G1_1}%
  \BibitemOpen
  \bibfield  {author} {\bibinfo {author} {\bibfnamefont {D.}~\bibnamefont
  {Feller}}, \bibinfo {author} {\bibfnamefont {K.~A.}\ \bibnamefont
  {Peterson}}, \ and\ \bibinfo {author} {\bibfnamefont {D.~A.}\ \bibnamefont
  {Dixon}},\ }\href {\doibase http://dx.doi.org/10.1063/1.3008061} {\bibfield
  {journal} {\bibinfo  {journal} {J. Chem. Phys.}\ }\textbf {\bibinfo {volume}
  {129}},\ \bibinfo {eid} {204105} (\bibinfo {year} {2008})}\BibitemShut
  {NoStop}%
\bibitem [{\citenamefont {Grossman}(2002)}]{G1_2}%
  \BibitemOpen
  \bibfield  {author} {\bibinfo {author} {\bibfnamefont {J.~C.}\ \bibnamefont
  {Grossman}},\ }\href {\doibase http://dx.doi.org/10.1063/1.1487829}
  {\bibfield  {journal} {\bibinfo  {journal} {J. Chem. Phys.}\ }\textbf
  {\bibinfo {volume} {117}},\ \bibinfo {pages} {1434} (\bibinfo {year}
  {2002})}\BibitemShut {NoStop}%
\bibitem [{\citenamefont {Becke}(1992)}]{G1_DFT}%
  \BibitemOpen
  \bibfield  {author} {\bibinfo {author} {\bibfnamefont {A.~D.}\ \bibnamefont
  {Becke}},\ }\href {\doibase http://dx.doi.org/10.1063/1.462066} {\bibfield
  {journal} {\bibinfo  {journal} {J. Chem. Phys.}\ }\textbf {\bibinfo {volume}
  {96}},\ \bibinfo {pages} {2155} (\bibinfo {year} {1992})}\BibitemShut
  {NoStop}%
\bibitem [{\citenamefont {Ernzerhof}\ and\ \citenamefont
  {Scuseria}(1999)}]{G1_DFT_1}%
  \BibitemOpen
  \bibfield  {author} {\bibinfo {author} {\bibfnamefont {M.}~\bibnamefont
  {Ernzerhof}}\ and\ \bibinfo {author} {\bibfnamefont {G.~E.}\ \bibnamefont
  {Scuseria}},\ }\href {\doibase http://dx.doi.org/10.1063/1.478401} {\bibfield
   {journal} {\bibinfo  {journal} {J. Chem. Phys.}\ }\textbf {\bibinfo {volume}
  {110}},\ \bibinfo {pages} {5029} (\bibinfo {year} {1999})}\BibitemShut
  {NoStop}%
\bibitem [{\citenamefont {Nemec}, \citenamefont {Towler},\ and\ \citenamefont
  {Needs}(2010)}]{G1_3}%
  \BibitemOpen
  \bibfield  {author} {\bibinfo {author} {\bibfnamefont {N.}~\bibnamefont
  {Nemec}}, \bibinfo {author} {\bibfnamefont {M.~D.}\ \bibnamefont {Towler}}, \
  and\ \bibinfo {author} {\bibfnamefont {R.~J.}\ \bibnamefont {Needs}},\ }\href
  {\doibase http://dx.doi.org/10.1063/1.3288054} {\bibfield  {journal}
  {\bibinfo  {journal} {J. Chem. Phys.}\ }\textbf {\bibinfo {volume} {132}},\
  \bibinfo {eid} {034111} (\bibinfo {year} {2010})}\BibitemShut {NoStop}%
\bibitem [{\citenamefont {Morales}\ \emph {et~al.}(2012)\citenamefont
  {Morales}, \citenamefont {McMinis}, \citenamefont {Clark}, \citenamefont
  {Kim},\ and\ \citenamefont {Scuseria}}]{G1_4}%
  \BibitemOpen
  \bibfield  {author} {\bibinfo {author} {\bibfnamefont {M.~A.}\ \bibnamefont
  {Morales}}, \bibinfo {author} {\bibfnamefont {J.}~\bibnamefont {McMinis}},
  \bibinfo {author} {\bibfnamefont {B.~K.}\ \bibnamefont {Clark}}, \bibinfo
  {author} {\bibfnamefont {J.}~\bibnamefont {Kim}}, \ and\ \bibinfo {author}
  {\bibfnamefont {G.~E.}\ \bibnamefont {Scuseria}},\ }\href@noop {} {\bibfield
  {journal} {\bibinfo  {journal} {J. Chem. Theor. Comput.}\ }\textbf {\bibinfo
  {volume} {8}},\ \bibinfo {pages} {2181} (\bibinfo {year} {2012})}\BibitemShut
  {NoStop}%
\bibitem [{\citenamefont {Flyvbjerg}\ and\ \citenamefont
  {Petersen}(1989)}]{Blocking}%
  \BibitemOpen
  \bibfield  {author} {\bibinfo {author} {\bibfnamefont {H.}~\bibnamefont
  {Flyvbjerg}}\ and\ \bibinfo {author} {\bibfnamefont {H.~G.}\ \bibnamefont
  {Petersen}},\ }\href {\doibase http://dx.doi.org/10.1063/1.457480} {\bibfield
   {journal} {\bibinfo  {journal} {J. Chem. Phys.}\ }\textbf {\bibinfo {volume}
  {91}},\ \bibinfo {pages} {461} (\bibinfo {year} {1989})}\BibitemShut
  {NoStop}%
\bibitem [{\citenamefont {Adler}, \citenamefont {Knizia},\ and\ \citenamefont
  {Werner}(2007)}]{molpro_CCSD_F12}%
  \BibitemOpen
  \bibfield  {author} {\bibinfo {author} {\bibfnamefont {T.~B.}\ \bibnamefont
  {Adler}}, \bibinfo {author} {\bibfnamefont {G.}~\bibnamefont {Knizia}}, \
  and\ \bibinfo {author} {\bibfnamefont {H.-J.}\ \bibnamefont {Werner}},\
  }\href@noop {} {\bibfield  {journal} {\bibinfo  {journal} {J. Chem. Phys.}\
  }\textbf {\bibinfo {volume} {127}},\ \bibinfo {eid} {221106} (\bibinfo {year}
  {2007})}\BibitemShut {NoStop}%
\bibitem [{\citenamefont {Halkier}\ \emph {et~al.}(1999)\citenamefont
  {Halkier}, \citenamefont {Helgaker}, \citenamefont {J\o{}rgensen},
  \citenamefont {Klopper},\ and\ \citenamefont {Olsen}}]{HF_extra}%
  \BibitemOpen
  \bibfield  {author} {\bibinfo {author} {\bibfnamefont {A.}~\bibnamefont
  {Halkier}}, \bibinfo {author} {\bibfnamefont {T.}~\bibnamefont {Helgaker}},
  \bibinfo {author} {\bibfnamefont {P.}~\bibnamefont {J\o{}rgensen}}, \bibinfo
  {author} {\bibfnamefont {W.}~\bibnamefont {Klopper}}, \ and\ \bibinfo
  {author} {\bibfnamefont {J.}~\bibnamefont {Olsen}},\ }\href {\doibase
  http://dx.doi.org/10.1016/S0009-2614(99)00179-7} {\bibfield  {journal}
  {\bibinfo  {journal} {Chem. Phys. Lett.}\ }\textbf {\bibinfo {volume}
  {302}},\ \bibinfo {pages} {437 } (\bibinfo {year} {1999})}\BibitemShut
  {NoStop}%
\bibitem [{\citenamefont {Kong}\ and\ \citenamefont
  {Valeev}(2010)}]{Kong_2010}%
  \BibitemOpen
  \bibfield  {author} {\bibinfo {author} {\bibfnamefont {L.}~\bibnamefont
  {Kong}}\ and\ \bibinfo {author} {\bibfnamefont {E.~F.}\ \bibnamefont
  {Valeev}},\ }\href@noop {} {\bibfield  {journal} {\bibinfo  {journal} {J.
  Chem. Phys.}\ }\textbf {\bibinfo {volume} {133}},\ \bibinfo {eid} {174126}
  (\bibinfo {year} {2010})}\BibitemShut {NoStop}%
\bibitem [{\citenamefont {Werner}\ \emph {et~al.}(2012)\citenamefont {Werner},
  \citenamefont {Knowles}, \citenamefont {Knizia}, \citenamefont {Manby},\ and\
  \citenamefont {Sch{\"u}tz}}]{MOLPRO}%
  \BibitemOpen
  \bibfield  {author} {\bibinfo {author} {\bibfnamefont {H.}~\bibnamefont
  {Werner}}, \bibinfo {author} {\bibfnamefont {P.}~\bibnamefont {Knowles}},
  \bibinfo {author} {\bibfnamefont {G.}~\bibnamefont {Knizia}}, \bibinfo
  {author} {\bibfnamefont {F.}~\bibnamefont {Manby}}, \ and\ \bibinfo {author}
  {\bibfnamefont {M.}~\bibnamefont {Sch{\"u}tz}},\ }\href@noop {} {\enquote
  {\bibinfo {title} {{MOLPRO}, version 2012.1, a package of ab initio
  programs},}\ } (\bibinfo {year} {2012})\BibitemShut {NoStop}%
\bibitem [{\citenamefont {Knizia}, \citenamefont {Adler},\ and\ \citenamefont
  {Werner}(2009)}]{molpro_CCSD_F12_2}%
  \BibitemOpen
  \bibfield  {author} {\bibinfo {author} {\bibfnamefont {G.}~\bibnamefont
  {Knizia}}, \bibinfo {author} {\bibfnamefont {T.~B.}\ \bibnamefont {Adler}}, \
  and\ \bibinfo {author} {\bibfnamefont {H.-J.}\ \bibnamefont {Werner}},\
  }\href@noop {} {\bibfield  {journal} {\bibinfo  {journal} {J. Chem. Phys.}\
  }\textbf {\bibinfo {volume} {130}},\ \bibinfo {eid} {054104} (\bibinfo {year}
  {2009})}\BibitemShut {NoStop}%
\bibitem [{\citenamefont {Chase~{J}r}\ and\ \citenamefont
  {Tables}(1998)}]{G1_b31}%
  \BibitemOpen
  \bibfield  {author} {\bibinfo {author} {\bibfnamefont {M.}~\bibnamefont
  {Chase~{J}r}}\ and\ \bibinfo {author} {\bibfnamefont {N.-J.~T.}\ \bibnamefont
  {Tables}},\ }\href@noop {} {\bibfield  {journal} {\bibinfo  {journal} {J.
  Phys. Chem. Ref. Data}\ ,\ \bibinfo {pages} {1}} (\bibinfo {year}
  {1998})}\BibitemShut {NoStop}%
\bibitem [{\citenamefont {Huber}\ and\ \citenamefont
  {Herzberg}(1979)}]{G1_b32}%
  \BibitemOpen
  \bibfield  {author} {\bibinfo {author} {\bibfnamefont {K.}~\bibnamefont
  {Huber}}\ and\ \bibinfo {author} {\bibfnamefont {G.}~\bibnamefont
  {Herzberg}},\ }\href@noop {} {\emph {\bibinfo {title} {Constants of Diatomic
  Molecules, {V}ol. 4, Molecular Spectra and Molecular Structure}}}\ (\bibinfo
  {publisher} {Van Nostrand Reinhold, New York},\ \bibinfo {year}
  {1979})\BibitemShut {NoStop}%
\bibitem [{\citenamefont {Way}\ and\ \citenamefont {Stwalley}(1973)}]{G1_b33}%
  \BibitemOpen
  \bibfield  {author} {\bibinfo {author} {\bibfnamefont {K.~R.}\ \bibnamefont
  {Way}}\ and\ \bibinfo {author} {\bibfnamefont {W.~C.}\ \bibnamefont
  {Stwalley}},\ }\href@noop {} {\bibfield  {journal} {\bibinfo  {journal} {J.
  Chem. Phys.}\ }\textbf {\bibinfo {volume} {59}},\ \bibinfo {pages} {5298}
  (\bibinfo {year} {1973})}\BibitemShut {NoStop}%
\bibitem [{\citenamefont {Huang}, \citenamefont {Barts},\ and\ \citenamefont
  {Halpern}(1992)}]{G1_b34}%
  \BibitemOpen
  \bibfield  {author} {\bibinfo {author} {\bibfnamefont {Y.}~\bibnamefont
  {Huang}}, \bibinfo {author} {\bibfnamefont {S.~A.}\ \bibnamefont {Barts}}, \
  and\ \bibinfo {author} {\bibfnamefont {J.~B.}\ \bibnamefont {Halpern}},\
  }\href@noop {} {\bibfield  {journal} {\bibinfo  {journal} {J. Phys. Chem.}\
  }\textbf {\bibinfo {volume} {96}},\ \bibinfo {pages} {425} (\bibinfo {year}
  {1992})}\BibitemShut {NoStop}%
\bibitem [{\citenamefont {Lengel}\ and\ \citenamefont {Zare}(1978)}]{G1_b35}%
  \BibitemOpen
  \bibfield  {author} {\bibinfo {author} {\bibfnamefont {R.}~\bibnamefont
  {Lengel}}\ and\ \bibinfo {author} {\bibfnamefont {R.}~\bibnamefont {Zare}},\
  }\href@noop {} {\bibfield  {journal} {\bibinfo  {journal} {Journal of the
  American Chemical Society}\ }\textbf {\bibinfo {volume} {100}},\ \bibinfo
  {pages} {7495} (\bibinfo {year} {1978})}\BibitemShut {NoStop}%
\bibitem [{\citenamefont {McKellar}\ \emph {et~al.}(1983)\citenamefont
  {McKellar}, \citenamefont {Bunker}, \citenamefont {Sears}, \citenamefont
  {Evenson}, \citenamefont {Saykally},\ and\ \citenamefont
  {Langhoff}}]{G1_b36}%
  \BibitemOpen
  \bibfield  {author} {\bibinfo {author} {\bibfnamefont {A.}~\bibnamefont
  {McKellar}}, \bibinfo {author} {\bibfnamefont {P.}~\bibnamefont {Bunker}},
  \bibinfo {author} {\bibfnamefont {T.~J.}\ \bibnamefont {Sears}}, \bibinfo
  {author} {\bibfnamefont {K.}~\bibnamefont {Evenson}}, \bibinfo {author}
  {\bibfnamefont {R.~J.}\ \bibnamefont {Saykally}}, \ and\ \bibinfo {author}
  {\bibfnamefont {S.}~\bibnamefont {Langhoff}},\ }\href@noop {} {\bibfield
  {journal} {\bibinfo  {journal} {J. Chem. Phys.}\ }\textbf {\bibinfo {volume}
  {79}},\ \bibinfo {pages} {5251} (\bibinfo {year} {1983})}\BibitemShut
  {NoStop}%
\bibitem [{\citenamefont {Baulch}\ \emph {et~al.}(1982)\citenamefont {Baulch},
  \citenamefont {Cox}, \citenamefont {Crutzen}, \citenamefont {Hampson~Jr},
  \citenamefont {Kerr}, \citenamefont {Troe}, \citenamefont {Watson} \emph
  {et~al.}}]{G1_b37}%
  \BibitemOpen
  \bibfield  {author} {\bibinfo {author} {\bibfnamefont {D.}~\bibnamefont
  {Baulch}}, \bibinfo {author} {\bibfnamefont {R.}~\bibnamefont {Cox}},
  \bibinfo {author} {\bibfnamefont {P.}~\bibnamefont {Crutzen}}, \bibinfo
  {author} {\bibfnamefont {R.}~\bibnamefont {Hampson~Jr}}, \bibinfo {author}
  {\bibfnamefont {J.}~\bibnamefont {Kerr}}, \bibinfo {author} {\bibfnamefont
  {J.}~\bibnamefont {Troe}}, \bibinfo {author} {\bibfnamefont {R.}~\bibnamefont
  {Watson}},  \emph {et~al.},\ }\href@noop {} {\bibfield  {journal} {\bibinfo
  {journal} {Journal of Physical and Chemical Reference Data}\ }\textbf
  {\bibinfo {volume} {11}},\ \bibinfo {pages} {327} (\bibinfo {year}
  {1982})}\BibitemShut {NoStop}%
\bibitem [{\citenamefont {Gibson}, \citenamefont {Greene},\ and\ \citenamefont
  {Berkowitz}(1985)}]{G1_b38}%
  \BibitemOpen
  \bibfield  {author} {\bibinfo {author} {\bibfnamefont {S.}~\bibnamefont
  {Gibson}}, \bibinfo {author} {\bibfnamefont {J.}~\bibnamefont {Greene}}, \
  and\ \bibinfo {author} {\bibfnamefont {J.}~\bibnamefont {Berkowitz}},\
  }\href@noop {} {\bibfield  {journal} {\bibinfo  {journal} {J. Chem. Phys.}\
  }\textbf {\bibinfo {volume} {83}},\ \bibinfo {pages} {4319} (\bibinfo {year}
  {1985})}\BibitemShut {NoStop}%
\bibitem [{\citenamefont {Wagman}\ \emph {et~al.}(1982)\citenamefont {Wagman},
  \citenamefont {Evans}, \citenamefont {Parker}, \citenamefont {Schumm},
  \citenamefont {Halow}, \citenamefont {Bailey}, \citenamefont {Churney},\ and\
  \citenamefont {Nuttall}}]{G1_b39}%
  \BibitemOpen
  \bibfield  {author} {\bibinfo {author} {\bibfnamefont {D.~D.}\ \bibnamefont
  {Wagman}}, \bibinfo {author} {\bibfnamefont {W.~H.}\ \bibnamefont {Evans}},
  \bibinfo {author} {\bibfnamefont {V.~B.}\ \bibnamefont {Parker}}, \bibinfo
  {author} {\bibfnamefont {R.~H.}\ \bibnamefont {Schumm}}, \bibinfo {author}
  {\bibfnamefont {I.}~\bibnamefont {Halow}}, \bibinfo {author} {\bibfnamefont
  {S.~M.}\ \bibnamefont {Bailey}}, \bibinfo {author} {\bibfnamefont {K.~L.}\
  \bibnamefont {Churney}}, \ and\ \bibinfo {author} {\bibfnamefont {R.~L.}\
  \bibnamefont {Nuttall}},\ }\href@noop {} {\bibfield  {journal} {\bibinfo
  {journal} {J. Phys. Chem. Ref. Data, Suppl. 2}\ }\textbf {\bibinfo {volume}
  {11}} (\bibinfo {year} {1982})}\BibitemShut {NoStop}%
\bibitem [{\citenamefont {Lias}\ \emph {et~al.}(1988)\citenamefont {Lias},
  \citenamefont {Bartmess}, \citenamefont {Liebman}, \citenamefont {Holmes},
  \citenamefont {Levin},\ and\ \citenamefont {Mallard}}]{G1_b40}%
  \BibitemOpen
  \bibfield  {author} {\bibinfo {author} {\bibfnamefont {S.}~\bibnamefont
  {Lias}}, \bibinfo {author} {\bibfnamefont {J.}~\bibnamefont {Bartmess}},
  \bibinfo {author} {\bibfnamefont {J.}~\bibnamefont {Liebman}}, \bibinfo
  {author} {\bibfnamefont {J.}~\bibnamefont {Holmes}}, \bibinfo {author}
  {\bibfnamefont {R.}~\bibnamefont {Levin}}, \ and\ \bibinfo {author}
  {\bibfnamefont {W.~G.}\ \bibnamefont {Mallard}},\ }\href@noop {} {\emph
  {\bibinfo {title} {Ion Energetics Data}}}\ (\bibinfo  {publisher} {NIST
  Chemistry WebBook, NIST Standard Reference Database Number 69, Eds. P.J.
  Linstrom and W.G. Mallard, National Institute of Standards and Technology,
  Gaithersburg MD, 20899},\ \bibinfo {year} {1988})\BibitemShut {NoStop}%
\bibitem [{\citenamefont {Glushko}\ \emph {et~al.}(1978)\citenamefont
  {Glushko}, \citenamefont {Gurvich}, \citenamefont {Bergman}, \citenamefont
  {Veitz}, \citenamefont {Medvedev}, \citenamefont {Khachkuruzov},\ and\
  \citenamefont {Jungman}}]{G1_b42}%
  \BibitemOpen
  \bibfield  {author} {\bibinfo {author} {\bibfnamefont {V.~P.}\ \bibnamefont
  {Glushko}}, \bibinfo {author} {\bibfnamefont {L.~V.}\ \bibnamefont
  {Gurvich}}, \bibinfo {author} {\bibfnamefont {G.~A.}\ \bibnamefont
  {Bergman}}, \bibinfo {author} {\bibfnamefont {I.~V.}\ \bibnamefont {Veitz}},
  \bibinfo {author} {\bibfnamefont {V.~A.}\ \bibnamefont {Medvedev}}, \bibinfo
  {author} {\bibfnamefont {G.~A.}\ \bibnamefont {Khachkuruzov}}, \ and\
  \bibinfo {author} {\bibfnamefont {V.~S.}\ \bibnamefont {Jungman}},\
  }\href@noop {} {\emph {\bibinfo {title} {Thermodynamic properties of Pure
  Substances}}}\ (\bibinfo  {publisher} {Nauka, Moscow, USSR},\ \bibinfo {year}
  {1978})\BibitemShut {NoStop}%
\bibitem [{\citenamefont {Berkowitz}\ \emph {et~al.}(1987)\citenamefont
  {Berkowitz}, \citenamefont {Greene}, \citenamefont {Cho},\ and\ \citenamefont
  {Ruščić}}]{G1_b73}%
  \BibitemOpen
  \bibfield  {author} {\bibinfo {author} {\bibfnamefont {J.}~\bibnamefont
  {Berkowitz}}, \bibinfo {author} {\bibfnamefont {J.~P.}\ \bibnamefont
  {Greene}}, \bibinfo {author} {\bibfnamefont {H.}~\bibnamefont {Cho}}, \ and\
  \bibinfo {author} {\bibfnamefont {B.}~\bibnamefont {Ruščić}},\ }\href
  {\doibase http://dx.doi.org/10.1063/1.452213} {\bibfield  {journal} {\bibinfo
   {journal} {J. Chem. Phys.}\ }\textbf {\bibinfo {volume} {86}},\ \bibinfo
  {pages} {1235} (\bibinfo {year} {1987})}\BibitemShut {NoStop}%
\bibitem [{\citenamefont {Butler}, \citenamefont {Kawaguchi},\ and\
  \citenamefont {Hirota}(1983)}]{G1_b80}%
  \BibitemOpen
  \bibfield  {author} {\bibinfo {author} {\bibfnamefont {J.~E.}\ \bibnamefont
  {Butler}}, \bibinfo {author} {\bibfnamefont {K.}~\bibnamefont {Kawaguchi}}, \
  and\ \bibinfo {author} {\bibfnamefont {E.}~\bibnamefont {Hirota}},\
  }\href@noop {} {\bibfield  {journal} {\bibinfo  {journal} {J. Molec. Spect.}\
  }\textbf {\bibinfo {volume} {101}},\ \bibinfo {pages} {161} (\bibinfo {year}
  {1983})}\BibitemShut {NoStop}%
\bibitem [{\citenamefont {Berkowitz}\ \emph {et~al.}(1986)\citenamefont
  {Berkowitz}, \citenamefont {Curtiss}, \citenamefont {Gibson}, \citenamefont
  {Greene}, \citenamefont {Hillhouse},\ and\ \citenamefont {Pople}}]{G1_b81}%
  \BibitemOpen
  \bibfield  {author} {\bibinfo {author} {\bibfnamefont {J.}~\bibnamefont
  {Berkowitz}}, \bibinfo {author} {\bibfnamefont {L.}~\bibnamefont {Curtiss}},
  \bibinfo {author} {\bibfnamefont {S.}~\bibnamefont {Gibson}}, \bibinfo
  {author} {\bibfnamefont {J.}~\bibnamefont {Greene}}, \bibinfo {author}
  {\bibfnamefont {G.}~\bibnamefont {Hillhouse}}, \ and\ \bibinfo {author}
  {\bibfnamefont {J.}~\bibnamefont {Pople}},\ }\href@noop {} {\bibfield
  {journal} {\bibinfo  {journal} {J. Chem. Phys.}\ }\textbf {\bibinfo {volume}
  {84}},\ \bibinfo {pages} {375} (\bibinfo {year} {1986})}\BibitemShut
  {NoStop}%
\bibitem [{\citenamefont {Dattani}(2015)}]{Dattani2015}%
  \BibitemOpen
  \bibfield  {author} {\bibinfo {author} {\bibfnamefont {N.~S.}\ \bibnamefont
  {Dattani}},\ }\href {\doibase 10.1016/j.jms.2014.09.005} {\bibfield
  {journal} {\bibinfo  {journal} {J. Molec. Spect.}\ }\textbf {\bibinfo
  {volume} {311}},\ \bibinfo {pages} {76} (\bibinfo {year} {2015})}\BibitemShut
  {NoStop}%
\bibitem [{\citenamefont {{Le Roy}}\ \emph {et~al.}(2009)\citenamefont {{Le
  Roy}}, \citenamefont {Dattani}, \citenamefont {Coxon}, \citenamefont {Ross},
  \citenamefont {Crozet},\ and\ \citenamefont {Linton}}]{LeRoy2009}%
  \BibitemOpen
  \bibfield  {author} {\bibinfo {author} {\bibfnamefont {R.~J.}\ \bibnamefont
  {{Le Roy}}}, \bibinfo {author} {\bibfnamefont {N.~S.}\ \bibnamefont
  {Dattani}}, \bibinfo {author} {\bibfnamefont {J.~A.}\ \bibnamefont {Coxon}},
  \bibinfo {author} {\bibfnamefont {A.~J.}\ \bibnamefont {Ross}}, \bibinfo
  {author} {\bibfnamefont {P.}~\bibnamefont {Crozet}}, \ and\ \bibinfo {author}
  {\bibfnamefont {C.}~\bibnamefont {Linton}},\ }\href {\doibase
  10.1063/1.3264688} {\bibfield  {journal} {\bibinfo  {journal} {J. Chem.
  Phys.}\ }\textbf {\bibinfo {volume} {131}},\ \bibinfo {pages} {204309}
  (\bibinfo {year} {2009})}\BibitemShut {NoStop}%
\bibitem [{\citenamefont {Feller}\ and\ \citenamefont
  {Peterson}(1999)}]{feller_benchmark}%
  \BibitemOpen
  \bibfield  {author} {\bibinfo {author} {\bibfnamefont {D.}~\bibnamefont
  {Feller}}\ and\ \bibinfo {author} {\bibfnamefont {K.~A.}\ \bibnamefont
  {Peterson}},\ }\href@noop {} {\bibfield  {journal} {\bibinfo  {journal} {J.
  Chem. Phys.}\ }\textbf {\bibinfo {volume} {110}},\ \bibinfo {pages} {8384}
  (\bibinfo {year} {1999})}\BibitemShut {NoStop}%
\bibitem [{\citenamefont {Fr{\'e}chet}(1927)}]{Weibull1}%
  \BibitemOpen
  \bibfield  {author} {\bibinfo {author} {\bibfnamefont {M.}~\bibnamefont
  {Fr{\'e}chet}},\ }\bibfield  {booktitle} {\emph {\bibinfo {booktitle}
  {Annales de la societe Polonaise de Mathematique}},\ }\href@noop {} {\
  \textbf {\bibinfo {volume} {5}},\ \bibinfo {pages} {93} (\bibinfo {year}
  {1927})}\BibitemShut {NoStop}%
\bibitem [{\citenamefont {Weibull}(1951)}]{Weibull2}%
  \BibitemOpen
  \bibfield  {author} {\bibinfo {author} {\bibfnamefont {W.}~\bibnamefont
  {Weibull}},\ }\href@noop {} {\bibfield  {journal} {\bibinfo  {journal} {ASME
  Journal of Applied Mechanics}\ ,\ \bibinfo {pages} {293}} (\bibinfo {year}
  {1951})}\BibitemShut {NoStop}%
\bibitem [{\citenamefont {Gruneis}\ \emph {et~al.}(2013)\citenamefont
  {Gruneis}, \citenamefont {Shepherd}, \citenamefont {Alavi}, \citenamefont
  {Tew},\ and\ \citenamefont {Booth}}]{Booth_PlaneW}%
  \BibitemOpen
  \bibfield  {author} {\bibinfo {author} {\bibfnamefont {A.}~\bibnamefont
  {Gruneis}}, \bibinfo {author} {\bibfnamefont {J.~J.}\ \bibnamefont
  {Shepherd}}, \bibinfo {author} {\bibfnamefont {A.}~\bibnamefont {Alavi}},
  \bibinfo {author} {\bibfnamefont {P.~T.}\ \bibnamefont {Tew}}, \ and\
  \bibinfo {author} {\bibfnamefont {G.~H.}\ \bibnamefont {Booth}},\ }\href@noop
  {} {\bibfield  {journal} {\bibinfo  {journal} {J. Chem. Phys.}\ }\textbf
  {\bibinfo {volume} {139}},\ \bibinfo {pages} {084112} (\bibinfo {year}
  {2013})}\BibitemShut {NoStop}%
\bibitem [{\citenamefont {Gr\"uneis}(2015)}]{PhysRevLett.115.066402}%
  \BibitemOpen
  \bibfield  {author} {\bibinfo {author} {\bibfnamefont {A.}~\bibnamefont
  {Gr\"uneis}},\ }\href@noop {} {\bibfield  {journal} {\bibinfo  {journal}
  {Phys. Rev. Lett.}\ }\textbf {\bibinfo {volume} {115}},\ \bibinfo {pages}
  {066402} (\bibinfo {year} {2015})}\BibitemShut {NoStop}%
\bibitem [{\citenamefont {Usvyat}(2013)}]{Usvyat_2013}%
  \BibitemOpen
  \bibfield  {author} {\bibinfo {author} {\bibfnamefont {D.}~\bibnamefont
  {Usvyat}},\ }\href@noop {} {\bibfield  {journal} {\bibinfo  {journal} {J.
  Chem. Phys.}\ }\textbf {\bibinfo {volume} {139}},\ \bibinfo {eid} {194101}
  (\bibinfo {year} {2013})}\BibitemShut {NoStop}%
\end{thebibliography}
\end{document}